\documentclass[12pt, draftclsnofoot, onecolumn]{IEEEtran}
\usepackage[super]{nth}

\usepackage{cite}
\usepackage{amsmath,amssymb,amsfonts}
\usepackage{graphicx}
\usepackage{textcomp}
\usepackage{xcolor}
\usepackage{algorithmicx,algpseudocode}
\usepackage{algpseudocode,algorithm}
\usepackage{booktabs}

\newcommand\argmin{\mathop{\mbox{{\rm argmin}}}\limits}
\algrenewcommand\algorithmicrequire{\textbf{Precondition:}}
\algrenewcommand\algorithmicensure{\textbf{Postcondition:}}

\begin{document}
	
	\title{Markov Decision Policies for Dynamic Video Delivery in Wireless Caching Networks
	}
	
	\author{Minseok Choi,
		Albert No,
		Mingyue Ji~\IEEEmembership{Member,~IEEE,}
		and~\\Joongheon Kim,~\IEEEmembership{Senior Member,~IEEE}% <-this % stops a space
		\thanks{M. Choi and J. Kim are with the School of Computer Science and Engineering, Chung-Ang University, Seoul, Korea, e-mail: ejaqmf@kaist.ac.kr, joongheon@cau.ac.kr.}% <-this % stops a space
		\thanks{A. No is with the Department of Electronic and Electrical engineering, Hongik University, Seoul, Korea, e-mail: albertno@hongik.ac.kr.}
		\thanks{M. Ji is with the Department of Electrical and Computer Engineering, The University of Utah, UT, United States, e-mails: mingyue.ji@utah.edu.}% <-this % stops a space
	}
	
	\maketitle
	
	\begin{abstract}
		
		This paper proposes a video delivery strategy for dynamic streaming services which maximizes time-average streaming quality under a playback delay constraint in wireless caching networks.
		The network where popular videos encoded by scalable video coding are already stored in randomly distributed caching nodes is considered under adaptive video streaming concepts, and distance-based interference management is investigated in this paper. 
		In this network model, a streaming user makes delay-constrained decisions depending on stochastic network states: 1) caching node for video delivery, 2) video quality, and 3) the quantity of video chunks to receive.
		Since wireless link activation for video delivery may introduce delays, different timescales for updating caching node association, video quality adaptation, and chunk amounts are considered.
		After associating with a caching node for video delivery, the streaming user chooses combinations of quality and chunk amounts in the small timescale.
		The dynamic decision making process for video quality and chunk amounts at each slot is modeled using Markov decision process, and the caching node decision is made based on the framework of Lyapunov optimization.
		Our intensive simulations verify that the proposed video delivery algorithm works reliably and also can control the tradeoff between video quality and playback latency. 
		
	\end{abstract}
	
	
	\section{Introduction}
	Within few years, it has been expected that tens of exabytes of global data traffic be handled on daily basis, and on-demand video streaming will account for about 70\% of them \cite{cisco}.
	%supporting rapidly growing amounts of data traffic generated by massive number of wireless mobile devices is one of the most rewarding challenges for next generation wireless communications.
	In on-demand video streaming services, a relatively small number of popular contents is requested at ultra high rates and playback delay is one of the most important measurement criteria of goodness \cite{youtube,mm17koo}.
	To deal with the characteristics, wireless caching technologies have been studied for video streaming services by storing popular videos in caching helpers located nearby users during off-peak time~\cite{femtocaching,CM2014Bastug,CM2014Wang}.
	Therefore, it is obvious that storing and streaming of video files are of major research interests in wireless caching networks.
	
	There have been major research results for caching popular files in stochastic wireless caching networks~\cite{caching:ICC2015Blaszczyszyn,CL2017Chen,caching:TWC2016Chae,caching:TC2016Malak,JSAC2016Gregori}. 
	The major goal of those research results was to design the optimal caching policies according to the popularity distribution of contents and wireless network topology.
	The probabilistic caching policy was proposed in \cite{caching:ICC2015Blaszczyszyn} to adapt characteristics of the stochastic network.
	Many probabilistic caching methods have been proposed depending on various optimization goals, e.g., maximization of cache hit probability \cite{caching:ICC2015Blaszczyszyn}, cache-aided throughput \cite{CL2017Chen}, average success probability of content delivery \cite{caching:TWC2016Chae}, density of successful reception \cite{caching:TC2016Malak}, and average video quality \cite{caching:ICC2019Choi}. 
	The authors of \cite{JSAC2016Gregori} considered a joint optimization of caching and delivery when user demands were known in advance.
	In addition, the optimal caching policy which maximizes the cache hit probability in two-tier networks with opportunistic spectrum access was designed in \cite{TWC2018Emara}. 
	However, these works on the caching policy do not consider the identical content with different qualities.
	
	Since video files can be encoded to multiple versions which differ in the quality levels, the video caching policies having different quality levels have been widely studied in~\cite{cachingDiffQual:infocom2014Poularakis,cachingDiffQual:CL2017Zhan,infocom2016Poularakis,cachingDiffQual:CL2016Wu}.
	Many researchers have proposed the static content placement policies under the consideration of differentiated quality requests for the same content, given probabilistic quality requests \cite{cachingDiffQual:infocom2014Poularakis}, \cite{cachingDiffQual:CL2017Zhan} or minimum quality requirements \cite{infocom2016Poularakis}. 
	Further, the probabilistic caching policy for video files of various quality levels was presented in \cite{cachingDiffQual:CL2016Wu} by using stochastic geometry, given the user preference for quality level.
	The above works are focused only on the content placement problem with different qualities, however, the delivery policy of contents with different qualities has not yet been studied much. 
	
	For video delivery/streaming, there are some necessary decisions to be made: 1) which caching node will deliver the video, 2) which quality of video will be provided, and 3) how many video chunks will be transmitted.
	%For video delivery/streaming, the system has to decide which caching node will deliver the content requested from the user, and this problem is usually 
	The first one is %named to 
	called {\em node association problem}, and in most research contributions that do not consider different quality levels for the same file, the file-requesting user is allowed to receive the desired video from the caching node under the strongest channel condition \cite{caching:TWC2016Chae,TWC2016Yang}.
	The node associations for video delivery in heterogeneous caching networks have been studied in \cite{NA:TC2014Poularakis,NA:TC2016Zhang,NA:TMC2017Jiang}.
	On the other hand, when videos with different qualities are independently cached, more elaborate node association algorithm is necessary, because the node association is consistent with decision on the content quality. 
	In this case, the video delivery policy was proposed in \cite{JSAC2018Choi} to pursue time-average video quality maximization while avoiding playback delays.
	
	Since dynamic video streaming allows each chunk to have a different quality depending on time-varying network conditions, some researchers addressed transmission schemes which serve the video by dynamically selecting the quality level \cite{VD_DiffQ:TM2013Wang}. 
	In \cite{VD_DiffQ:TC2015Bethanabhotla} and \cite{VD_DiffQ:TON2016Kim}, the scheduling policies which maximize the network utility function of time-averaged video quality in small-cell networks and device-to-device networks were proposed.
	The authors of \cite{TC2018Yang} considered scalable video coding (SVC) and proposed dynamic resource allocation and quality selection under the pricing strategy for interference.
	While the video delivery policies of \cite{VD_DiffQ:TM2013Wang,VD_DiffQ:TC2015Bethanabhotla,VD_DiffQ:TON2016Kim,TC2018Yang} are operated at the base station (BS) side, however the decision policy of video quality level at user sides was not considered.
	This scenario is consistent with the practical real-world software implementation of dynamic adaptive streaming over HTTP (DASH) \cite{dash}, in which users dynamically choose the most appropriate video quality.
	Even though the work of \cite{JSAC2018Choi} can choose the video quality at the user side, however it cannot dynamically change the video quality without updates of node association.
	
	Further, control of the amount of receiving chunks depending on stochastic network states has been largely neglected in above existing researches about video delivery.
	Even though the authors of \cite{JSAC2018Choi} and \cite{TC2018Yang} maximize the long-term time-average video quality under the various constraints, their metrics representing video quality is obtained by averaging the number of quality selections at each time slot. 
	This method would be not enough to evaluate the user's quality of service, especially when the transmission rate varies over the video streaming service time.
	In practice, when channel experiences deep fading and only the low-quality video is available, it would not be the best choice to receive as many chunks as the channel condition can provide.
	Rather than receiving many low-quality chunks, the user could prefer to wait channel conditions to be better and then to receive high-quality chunks, if it guarantees no playback delay.
	Therefore, by considering decision process of combinations of video quality and chunk amounts, we can formulate the optimization problem which maximizes the average video quality per each received chunk.
	
	This paper proposes a video delivery policy in the wireless caching network for dynamic streaming services. 
	The main contributions are as follows:
	\begin{itemize}
		\item This paper proposes dynamic video delivery policy depending on stochastic network states. The proposed policy makes three different but necessary decisions for the streaming user: 1) caching node for video delivery, 2) video quality and 3) the quantity of video chunks to receive.
		To the best of the authors' knowledge, no research has yet considered all of those video delivery decisions.
		
		\item Caching node association and decisions of video quality and the amount of receiving chunks are conducted in different timescales. 
		Since wireless link activation for video delivery is time-consuming, it is reasonable that caching node association is performed slower than decisions of video quality and the amount of receiving chunks.
		
		\item The optimization framework of video delivery policy is constructed based on frame-based Lypunov optimization theory \cite{TAC2013Neely} and Markov decision process.
		The optimal caching node is found by Lyapunov optimization while decisions of video quality and the amount of receiving chunks are made by using dynamic programming \cite{DynamicProgram}. 
		
		\item The proposed technique maximizes the average streaming quality while averting playback latency, and can control the tradeoff between video quality and playback delay.
		Different from \cite{JSAC2018Choi} and \cite{TC2018Yang}, we adopt the long-term average video quality per each received chunk as a performance metric.
		
		\item We perform simulations to verify the proposed video delivery policy and to show the advantages of using Lyapunov optimization theory and Markov decision process.
	\end{itemize}
	
	The rest of the paper is organized as follows. 
	The wireless video caching network model is described in Sec.~\ref{sec:network_model}.
	The optimization problem for dynamic video delivery is formulated in Sec.~\ref{sec:prob_formulation}.
	The rule of caching node association and control policies of quality level and receiving chunk amounts are proposed in Sec.~\ref{sec:caching_node_decision} and Sec.~\ref{sec:qual_chunk_decisions}.
	Simulation results are presented in Sec.~\ref{sec:simulation} and Sec.~\ref{sec:conclusion} concludes this paper.
	
	\section{Network Model}
	\label{sec:network_model}
	
	\subsection{Wireless caching network model}
	
\begin{figure} [t!]
		\centering
		\includegraphics[width=0.75\textwidth]{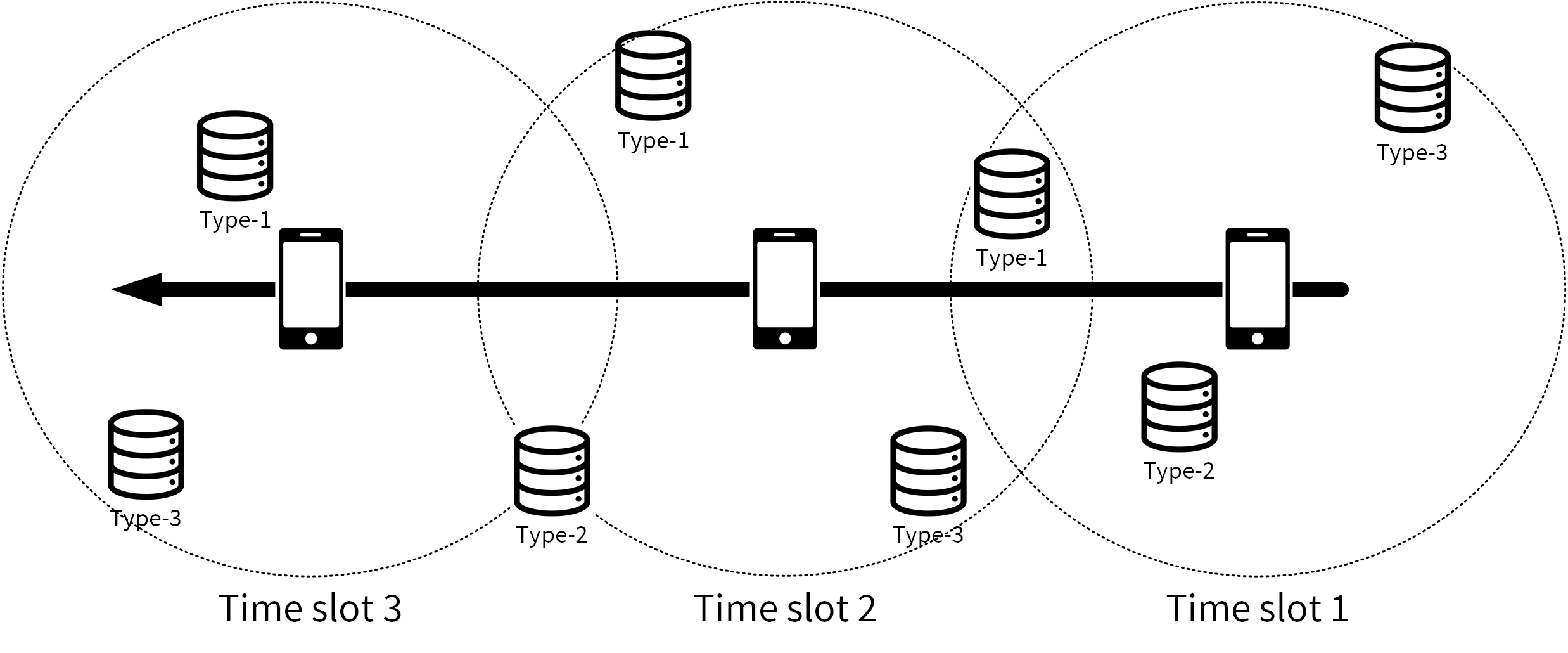}
		\caption{Network Model}
		\label{fig:network_model}
	\end{figure}

	This paper considers wireless caching network model where a user requests certain video file for one of caching nodes %within the radius of $R$, 
	around the user, as shown in Fig.~\ref{fig:network_model}. 
	The BS has already pushed popular video files during off-peak hours to caching nodes which have the finite storage size. %can store the finite number of videos. 
	Since we focus on video delivery, the caching policy is out of scope for this paper and only the desired video is considered. 
	Suppose that the desired video has $L$ quality levels. %and is encoded by layered encoding \cite{LayerEnc:Hartanto}. 
	Therefore, there are $L$ types of caching nodes, and the type-$l$ caching nodes can deliver the video of any quality $q \in \mathcal{L}_l$, where $\mathcal{L}_l=\{1, \cdots, l \}$ is the set of qualities which the type-$l$ caching node can provide.
	Thus, the type-$L$ caching nodes can provide all quality levels from the quality set $\mathcal{L}_L$.
	Note that simple definition of $\mathcal{L}_l=\{1, \cdots, l \}$ is assumed, but the proposed technique can be coordinated with any arbitrary quality set as long as multiple versions of the same video having different qualities are stored in caching nodes.
	
	The identical files of different qualities are stored in multiple caching nodes, and the type-$l$ caching nodes are distributed by the independent Poisson Point Processes (PPPs) with intensity $\lambda p_q$ \cite{caching:ICC2015Blaszczyszyn}, where $p_q$ indicates the caching probability of the requested video encoded to provide any quality $q \in \{1,\cdots,l\}$.
	Suppose that the caching policy is already determined, i.e., all $p_q$ for all $q$ are given.
	In addition, videos of different qualities have different file sizes and $N_q$ denotes the file size of quality $q$ in bits, satisfying $N_m < N_q$ for all $m, q \in \mathcal{L}_L$ and $m < q$.
	
	User mobility is also captured in network model.
	The user is moving towards certain direction and periodically searches for a caching node to receive the desired video file. %within the radius $R$.
	As shown in Fig.~\ref{fig:network_model}, geological distribution of caching nodes around the user varies at each time slot, so the caching node decision should be appropriately updated.
	Further, this paper also considers how many chunks of which quality level to be requested from the user depending on the stochastic network environment.
	When there are other users who exploit the wireless caching network with the same resource, the target streaming user is interfering with them.
	We adopt the distance-based interference management to limit the interference power lower than certain threshold, and details are explained in Section \ref{subsec:interference}.
	
	\subsection{User queue model and channel model}
	
	A video file consists of many sequential chunks. 
	The user receives the video file from a caching node and processes data for the streaming service in units of chunks. 
	Each chunk of a file is responsible for some playback time of the entire stream. 
	As long as all chunks are in correct sequence, each chunk can have different quality in dynamic streaming. 
	Therefore, the user can dynamically choose video quality level in each chunk processing time. 
	By using the queueing model, it can be said that the playback delay occurs when the queue does not have the chunk to be played.
	%the chunk to be played does not yet arrive at the queue.
	In this sense, receiver queue dynamics collectively reflects the various factors which cause the playback delay.
	
	In general, the user model has its own arrival and departure processes.
	The user queue dynamics in each discrete time slot $t \in \{0,1,\cdots \}$ can be represented as follows:
	\begin{equation}
	Q(t+1) = \max \{ Q(t) - c, 0 \} + M(t) \text{ and } Q(0)=0, \label{eq:q_dynamics1}
	\end{equation}
	where $Q(t)$ stands for the queue backlog at time $t$. 
	In addition, the departure $c$ is a constant because the streaming user does not change the video playback rate in general. 
	The arrival $M(t)$ denotes the number of received chunks at time $t$.
	
	Let the caching node which the user chooses for video delivery be $\alpha$.
	Then, $h(\alpha,t)=\sqrt{D(\alpha, t)}u(t)$ represents the Rayleigh fading channel between the user and the caching node $\alpha$ at time $t$, where $D(\alpha, t)=1/d(\alpha, t)^2$ controls path loss with $d(\alpha, t)$ being the user-caching node distance at time $t$ and $u$ represents the fast fading component having a complex Gaussian distribution, $u\sim \mathcal{C}\mathcal{N}(0,1)$.
	The link rate can be simply given by 
	$R(\alpha,t) = \mathcal{W} \log_2 \Big( 1+ \frac{\Psi|h(\alpha,t)|^2}{\Upsilon + 1} \Big)$,
	where $\mathcal{W}$, $\Psi$, and $\Upsilon$ are bandwidth, transmit SNR, and interference-noise-ratio (INR), respectively.
	
	The number of received chunks necessarily depends on the caching node decision $\alpha$ and its link rate.
	In this paper, each slot interval is determined to be channel coherence time $t_c$.
	Then, the number of received chunks $M(t)$ is constrained by 
	\begin{equation}
	M(t) N_{q(t)} \leq t_c R(\alpha,t). 
	\end{equation}
	%\mj{where $t \in \{0, t_c, 2t_c, \cdots \}$.} 
	Since $M(t)$ and $N_{q(t)}$ are nonnegative integers, %it also satisfies 
	\begin{equation}
	M(t) N_{q(t)} \leq B(\alpha,t) = \lfloor t_c R(\alpha,t) \rfloor. \label{eq:chunk_const}
	\end{equation}
	Therefore, the decision of $M(t)$ depends on the decisions of $\alpha(t)$ and $q(t)$ and the random network event $R(\alpha, t)$.
	
	\subsection{Distance-based interference management}
	\label{subsec:interference}
	
	Although many existing works have investigated %effects of interference on the wireless caching network 
	complex interference management schemes such as interference alignment and interference cancellation, most of researches on the wireless caching and delivery policy have still %ignored the interference network, 
	used simple interference avoidance based interference management schemes, e.g., by spectrum sharing \cite{TWC2017Chen} or assuming the protocol model \cite{TWC2018Qin}. 
	For simplicity, this paper considers the distance-based interference control for node association (i.e., link activation) for video delivery. 
	The design ideas can be extended to other more sophisticated interference management schemes \cite{DySPAN2005Etkin, ICC2009Blomer}.
	
\begin{figure} [t!]
		\centering
		\includegraphics[width=0.6\columnwidth]{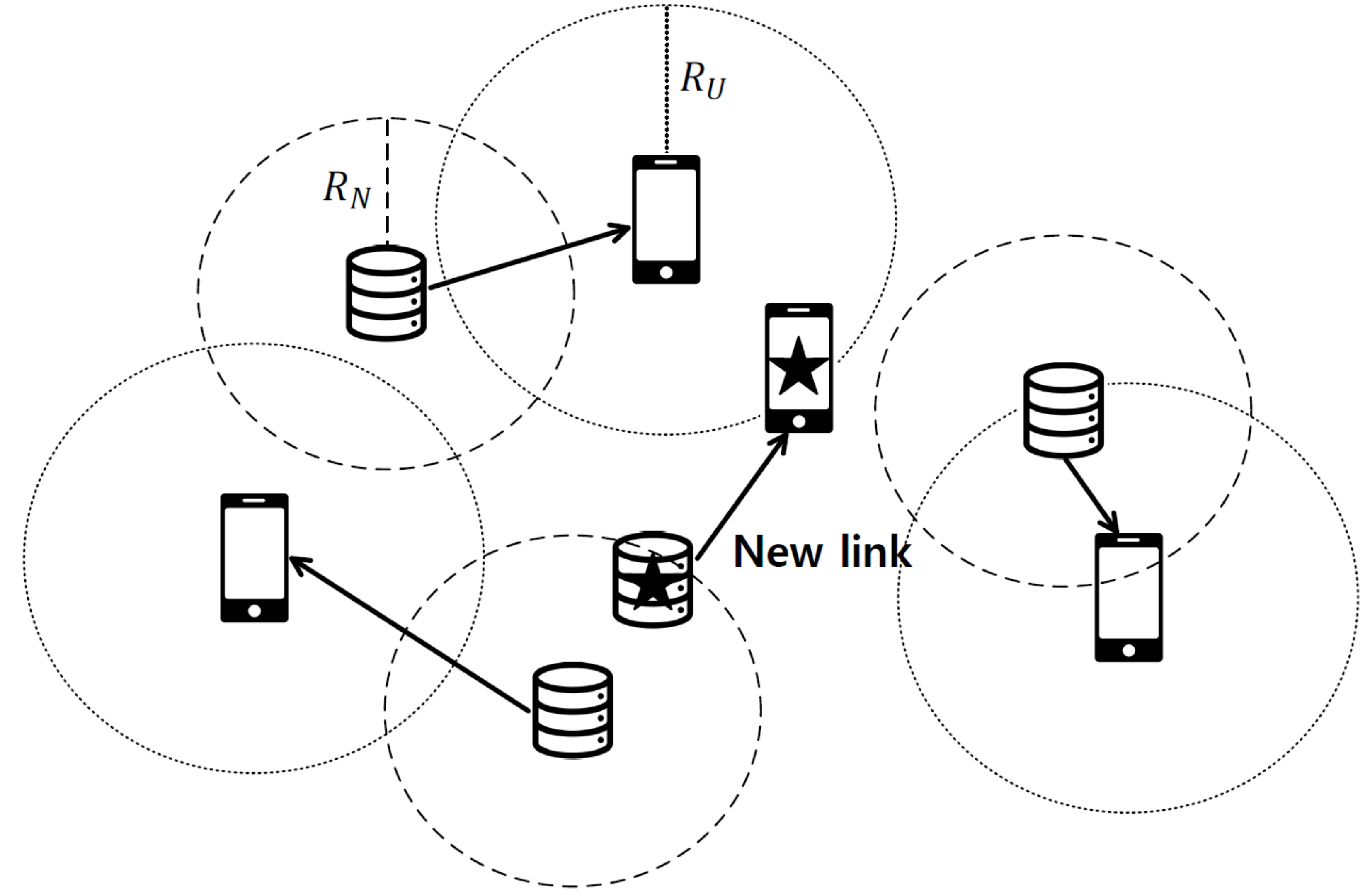}
		\caption{Safety radius and activation of new link for video delivery}
		\label{fig:safety_radii}
	\end{figure}

	Activation of the new link for video delivery in the wireless caching network means that the network allows the new streaming user to interfere with existing users.
	A new user causes two types of interference, 1) from the caching nodes already %delivering the video to 
	serving existing users to the new user, and 2) from the caching node associated with the new user to existing users.
	Therefore, we define $R_U$ and $R_N$ as the safety distances for streaming users and their associated caching nodes respectively to keep the interference levels below the predetermined threshold of $\rho$.
	In other words, a new streaming user who wants to exploit the wireless caching network should be generated outside the radius $R_N$ of all caching nodes associated with the existing users.
	In addition, the new user has to find the caching node to receive the desired content outside the radius $R_U$ of all existing users.
	The safety distances of $R_U$ and $R_N$ should be carefully chosen, and then %to limit interference power at both existing and the new users lower than threshold $\rho$. 
	%Therefore, we define the safety distances of $R_N$ and $R_U$ respectively for caching nodes and streaming users who already exist, to make INRs at every user lower than the threshold of $\rho$.
	a new pair of a caching node and a user can be generated only when their interference power is acceptable for every existing video delivery link, as shown in Fig. \ref{fig:safety_radii}.
	
	In this regard, a new pair of a caching node and a streaming user is allowed for video delivery through following two steps.
	The first step is to confirm the INR, say $\Upsilon_0$, at the new streaming user to be lower than $\rho$. 
	Here, $\Upsilon_0$ is the ratio of the aggregated interference power from all the activated caching nodes to noise variance. 
	If $\Upsilon_0 > \rho$, the system does not allow the new user to exploit the wireless caching network, and the new user should directly request the desired content from the server which has a whole file library or wait for content delivery in future. 
	For example, suppose that interference power from the nearest interfering caching node to the new user dominates $\Upsilon_0$.
	We further let $\Psi_0$ be the transmit SNR of the interfering node, and $d_n$ be the distance from the interfering node to the new user.
	%Suppose that interference power from the nearest interfering caching node to the new user dominates $\Upsilon_0$ and for example, denote $\Psi_0$ and $d_n$ by the transmit SNR of the interfering node and the distance from the interfering node to the new user, respectively.
	Then, INR becomes $\Upsilon_0 = \frac{\Psi_0}{d_n^2}$, and $R_N \geq \sqrt{\Psi_0/\rho}$ to guarantee $\Upsilon_0 \leq \rho$.
	
	Although the interference power at the new user is safe to exploit the wireless caching network, i.e., $\Upsilon_0 \leq \rho$, the caching node associated with the new user will be able to degrade the signal-to-interference-plus-noise ratios (SINRs) of existing users.
	Thus, the second step is required, in which the new user should find the caching node to receive the desired content with sufficiently large link rate as well as not to significantly interfere with other users. 
	Let one of existing users have a margin of INR to guarantee $\rho$ before activating the new link, denoted by $\delta = \rho - \Upsilon_0$.
	Since the interference signal from the new caching node is independent on other interfering nodes, $R_U = \sqrt{\Psi_0/\delta}$ is obtained similar to the case of $R_N$.
	Therefore, the new caching node should be chosen outside the radius of $R_U$ from every existing user whose margin of INR would be different from each other.
	
	Even though the new user and its caching node are found while limiting all interference levels at users lower than $\rho$, the newly generated link between them could not be enough to provide reliable content transmissions due to bad channel conditions.
	Therefore, we investigate the existence of the caching node around the new user which stores the requested content and can deliver the content reliably.
	%Therefore, we investigate whether or not at least one caching node which stores the content requested by the new user as well as enables to deliver the content reliably exists around the user.
	Let the minimum SINR for reliable video delivery denoted by $\gamma_{\text{min}}$.
	Then, the probability that at least one caching node can successfully deliver the desired content to the new user is represented by
	\begin{align}
	\eta = \mathrm{Pr} \bigg\{ \frac{\Psi|h_{n,1}|^2}{\Upsilon + 1} \geq \gamma_{\text{min}} \bigg\},
	\label{eq:eta}
	\end{align}
	where $h_{n,1}$ is the channel gain between the new user and the caching node whose channel condition is the strongest among the nodes storing the desired content of the user.
	According to order statistics and \cite{caching:TWC2016Chae}, the cumulative distribution function of the smallest reciprocal of channel power is $F_{\xi_{n,1}}(\xi) = 1-e^{-\pi\Gamma(2) \lambda_n \xi }$, where $\xi_{n,1}=1/|h_{n,1}|^2$ and $\lambda_n$ is the intensity of PPP of nodes caching the desired content.
	According to \eqref{eq:eta}, $\eta$ can be found by
	\begin{equation}
	\eta = 1 - \exp \bigg\{ -\pi \Gamma(2) \lambda_n \frac{\Psi}{\gamma_{\text{min}} (\Upsilon+1)} \bigg\}.
	\end{equation}
	
	Then, by introducing the minimum probability of finding at least one caching node for reliable video delivery denoted by $\eta_{\text{min}}$, a set of $\{\gamma_{\text{min}}, \eta_{\text{min}} \}$ can be considered as a criterion for new reliable link activation which satisfies $\eta\geq \eta_{\text{min}}$.
	In this regard, we can verify how much interference power is acceptable to satisfy the criterion of $\{\gamma_{\text{min}}, \eta_{\text{min}} \}$, as follows:
	\begin{align}
	&1 - \exp \bigg\{ -\pi \Gamma(2) \lambda_n \frac{\Psi}{\gamma_{\text{min}} (\Upsilon+1)} \bigg\} \geq \eta_{\text{min}} \nonumber \\
	&~\iff \frac{\pi \Gamma(2) \lambda_n \Psi}{\gamma_{\text{min}} \ln (\frac{1}{1-\eta_{\text{min}}})} - 1 \geq \Upsilon.
	\end{align}
	Thus, if all network parameters are given, the threshold of interference power can be determined by
	\begin{equation}
	\rho = \frac{\pi \Gamma(2) \lambda_n \Psi}{\gamma_{\text{min}} \ln (\frac{1}{1-\eta_{\text{min}}})} - 1.
	\end{equation}
	On the other hand, if the network requires the target criterion of interference management, i.e., $\rho$, $\gamma_{\text{min}}$, and $\eta_{\text{min}}$ are given, the system can determine how much transmit power is required and/or how many caching nodes store the desired video.
	
	In this paper, the minimum SINR threshold is set so that the chunk of the smallest size (i.e., the lowest quality) is deliverable at least,
	%is set to enable to delver a chunk of the smallest size (i.e., the lowest quality) at least, 
	i.e., $t_0 \mathcal{W}\log_2(1+\gamma_{\text{min}}) = N_1$.
	Then, we can say that caching nodes which store the desired content should be distributed with the intensity of $\lambda_{\text{min}}$ at least, as follows:
	\begin{equation}
	\lambda_n \geq \lambda_{\text{min}} = \frac{ (2^{\frac{N_1}{t_0\mathcal{W}}}-1) \ln (\frac{1}{1-\eta_{\text{min}}}) (1+\Upsilon) }{ \pi \Gamma(2) \Psi }.
	\end{equation}

	\section{Dynamic Video Delivery Policies}
	\label{sec:prob_formulation}
	
	\subsection{Video delivery decisions}
	
	The goal of this paper is to find the appropriate three decisions at each slot $t$ in the network model illustrated in Section \ref{sec:network_model}: 1) caching node for video delivery $\alpha(t)$, 2) video quality level $q(t)$, and 3) the quantity of receiving chunks $M(t)$.
	However, to update the caching node association, the time-consuming process is required in which the user sends the request signal for video delivery and the caching node approves it.
	Therefore, new caching node association is hardly performed as frequent as receiving chunks, and we suppose that the decision on $\alpha(t)$ is made at larger timescale than decisions on $q(t)$ and $M(t)$.
	
	\begin{figure} [h!]
		\centering
		\includegraphics[width=0.75\textwidth]{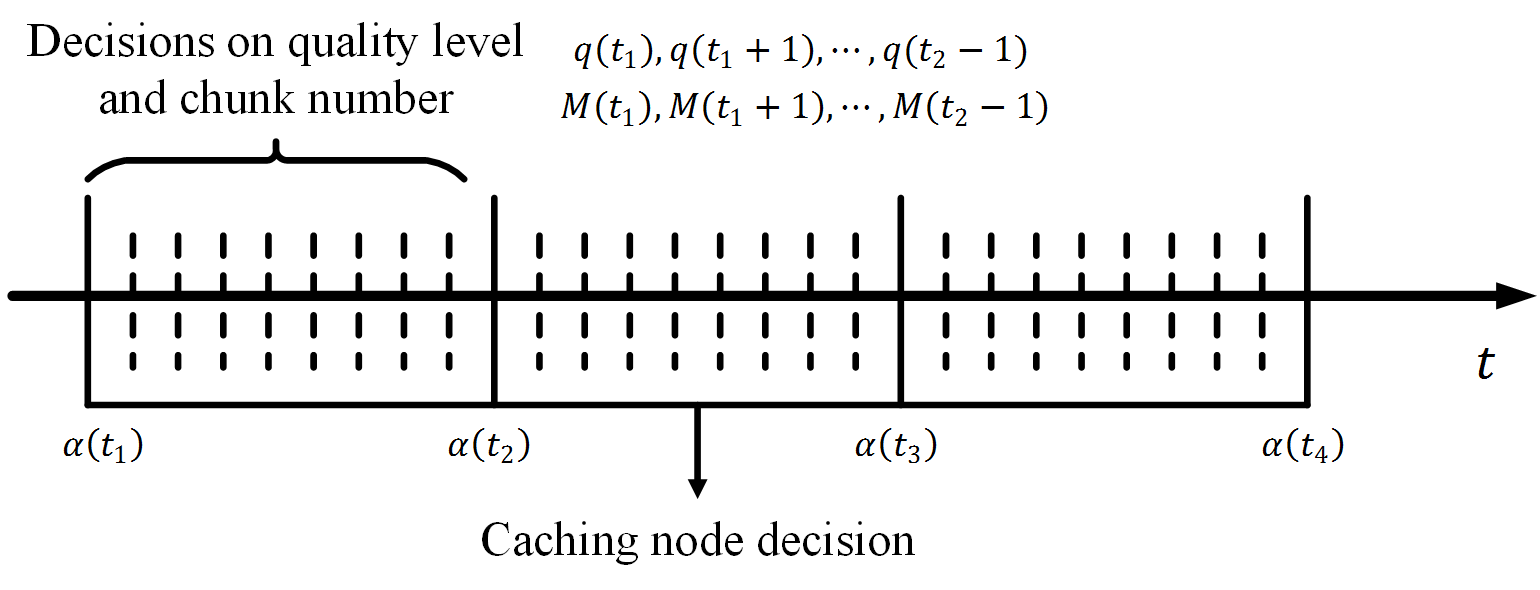}
		\caption{Different timescales for decisions on $\alpha(t)$, $q(t)$ and $M(t)$}
		\label{fig:decision_timescale}
	\end{figure}
	
	In this sense, the user decides $q(t)$ and $M(t)$ at time slots $t\in \{0,1,2\cdots \}$, but caching node decisions are performed at time slots $t=0,T,2T,\cdots$, where $T$ is the time interval for caching node association.
	The time slot for the $k$-th caching node decision is denoted by $t_k = (k-1)T$ for $k\in \{1,2,\cdots \}$. 
	Different timescales of decisions on $\alpha(t)$, $q(t)$ and $M(t)$ are described in Fig.~\ref{fig:decision_timescale}.
	Let the $k$-th frame for caching node decision be $\mathcal{T}_k = \{t_k,t_k+1,\cdots,t_k+T-1 \}$.
	As shown in Fig.~\ref{fig:decision_timescale}, after associating with the caching node $\alpha(t_k)$ at time $t_k$, decisions on quality level $q(t)$ and chunk amounts $M(t)$ are performed over $t\in \mathcal{T}_k$ to receive the desired video from $\alpha(t_k)$.
	Therefore, $q(t) \in \mathcal{L}_{l(\alpha(t_k))}$ 
	and $M(t) N_{q(t)} \leq B(\alpha(t_k),t)$ should be satisfied for $t\in \mathcal{T}_k$, where $l(\alpha(t_k))$ is the type of the caching node $\alpha(t_k)$.
	
	The user can make the candidate set of caching nodes denoted by $\mathcal{A}(t_k)$, and $\alpha(t_k)\in \mathcal{A}(t_k)$.
	All caching nodes in $\mathcal{A}(t_k)$ should be outside the radius $R_U$ of all existing users to limit the interference power lower than $\rho$.
	To avoid the situation in which no caching node can deliver the desired video, i.e., $\mathcal{A}(t_k) > 0$, the caching nodes which provide SINRs larger than $\gamma_{\text{min}}$ are assumed to be outside the radius $R_U$ of all existing users.
	$\mathcal{A}(t_k)$ consists of up to $L$ caching nodes, i.e., $|\mathcal{A}(t_k)| \leq L$, 
	in which each caching node in $\mathcal{A}(t_k)$ belongs to different types. 
	If there are several nodes of type-$l$, the user takes one of them whose channel condition is the strongest.
	There is no reason to choose another type-$l$ caching node while leaving the node with the strongest channel if another streaming user does not request the video from that strongest node. 
	In addition, $|\mathcal{A}(t_k)| < L$ means that $L-|\mathcal{A}(t_k)|$ caching node types do not exist around the user. 
	Suppose that the new streaming user is already generated outside the radius $R_N$ of all existing caching nodes and the INR $\Upsilon$ is observed at the new user. 
	Also, another user's link activation is banned around the target user due to the interference issue. 
	Then, we just consider the node association problem of the new streaming user with respect to the candidate set $\mathcal{A}(t_k)$ while the INR $\Upsilon$ is observed. 
	%Suppose that the new streaming user is also generated outside the radius $R_N$ of all existing caching nodes. 
	%Then, we just consider a streaming user in this network model and another user's link activation is banned around the target user due to the interference issue. 
	
	\subsection{Problem formulation}
	
	For determining the appropriate video delivery policy, two performance metrics are considered: playback delay and average streaming quality.
	Based on these goals, we can formulate the optimization problem which minimizes the quality degradation constrained on averting queue emptiness as follows:
	\begin{align}
	&\{ \boldsymbol{\alpha},\boldsymbol{q},\boldsymbol{M} \} = \underset{\alpha \in \mathcal{A},~q \in \mathcal{L}_{l(\alpha)} }{\argmin}~ \lim_{K\rightarrow \infty} \mathbb{E}\left[ \frac{1}{KT} \sum_{t=0}^{KT-1}(\bar{\mathcal{P}} - \mathcal{P}(q(t))) \cdot M(t) \right] \label{eq:opt_metric} \\
	&~~~~~~~~~~~~~~\text{s.t.}~\underset{t'\rightarrow \infty}{\lim} \frac{1}{t'} \sum_{t=0}^{t'-1}\nolimits \mathbb{E}[Z(t)] < \infty \label{eq:opt_const_queue} \\
	&~~~~~~~~~~~~~~~~~~~M(t)N_{q(t)} \leq B(\alpha,t) \label{eq:opt_const_chunk}
	\end{align}
	where $\mathcal{P}(q(t))$ is quality measure of $q(t)$ and $\bar{\mathcal{P}}$ is the maximum quality measure, i.e., equation \eqref{eq:opt_metric} is the time averaged video quality degradation.
	Decision vectors are represented as $\boldsymbol{\alpha}=[\alpha(t_1), \cdots, \alpha(t_K) ]$, $\boldsymbol{q}=[q(0),q(1),\cdots, q(KT-1)]$ and $\boldsymbol{M}=[M(0),M(1),\cdots,M(KT-1)]$.
	%Note \eqref{eq:opt_const_chunk} is consistent with \eqref{eq:chunk_const}.
	Specifically, the expectation of \eqref{eq:opt_metric} is with respect to random channel realizations and stochastic distributions of caching nodes.

	As mentioned earlier, playback delay occurs when the next chunk is not arrived in the queue, therefore the constraint \eqref{eq:opt_const_queue} has a role of avoiding queue emptiness, where $Z(t) = \tilde{Q} - Q(t)$.
	Here, $Z(t)$ is introduced to make $Q(t)$ large enough to avert playback delay, and $\tilde{Q}$ is a sufficiently large parameter which affects the maximal queue backlog. 
	From \eqref{eq:q_dynamics1}, the queue dynamics of $Z(t)$ can be represented as follows:
	\begin{equation}
	Z(t+1) = \min \{ Z(t) + c, \tilde{Q} \} - M(t) \text{ and } 
	Z(0) = \tilde{Q}.
	\end{equation}
	
	Even though the update rules of $Q(t)$ and $Z(t)$ are different, both queue dynamics mean the same video chunk processing. 
	Therefore, playback delay due to emptiness of $Q(t)$ can be explained by queueing delay of $Z(t)$. 
	By Littles' Law \cite{LittlesThm}, the expected value of $Z(t)$ is proportional to the time-averaged queueing delay.
	We aim to limit the queuing delay by addressing \eqref{eq:opt_const_queue}, and it is well known that Lyapunov optimization with \eqref{eq:opt_const_queue} can make $Z(t)$ bounded \cite{Lyapunov}.
	
	From the optimization problem \eqref{eq:opt_metric}-\eqref{eq:opt_const_chunk}, we can intuitively see how decisions are made depending on $Q(t)$. %and stochastic network environments.
	Suppose that the queue is almost empty. 
	In this case, the user prefers the caching node whose channel condition is strong, pursues low-quality file, and tries to receive as many chunks as possible to stack many chunks in the queue. 
	However, all of those decisions could degrade the average streaming quality. 
	When the caching node with the strongest channel condition belongs to type-$1$, it can be better to associate with the caching node of another type in terms of average quality. 
	In addition, when low quality is chosen, receiving too many chunks may not be a good choice. 
	The user would prefer to receive the small number of chunks in current time-step and wait the better channel condition.
	If the channel condition is improved at the next time-step, the user can request many chunks of high-quality video. 
	Thus, those decisions are strongly dependent on the queue state $Q(t)$, the caching node distribution, and channel conditions of caching node candidates.
	
	\section{Caching Node Decision Policy}
	\label{sec:caching_node_decision}
	
	For avoiding the queue emptiness, i.e., pursuing queue stability of $Z(t)$, the optimization problem of \eqref{eq:opt_metric}-\eqref{eq:opt_const_chunk} are solved based on the Lyapunov optimization theory.
	However, since the timescale of decision on $\boldsymbol{\alpha}$ is larger than that of decisions on $\boldsymbol{q}$ and $\boldsymbol{M}$, the frame-based Lyapunov optimization theory \cite{TAC2013Neely} is used for caching node decision.
	Lyapunov function $L(t)$ can be defined as 
	$L(t) = \frac{1}{2} Z(t)^2$.
	Then, let $\Delta(.)$ be a frame-based conditional Lyapunov function that can be formulated as $\Delta(t_k) = \mathbb{E}[L(t_k+T)-L(t_k)|Z(t_k)]$, i.e., the drift over the time interval $T$. 
	The dynamic policy is designed to solve the given optimization problem of \eqref{eq:opt_metric}-\eqref{eq:opt_const_chunk} by observing the current queue state, $Z(t_k)$, and determining the caching node to minimize a upper bound on frame-based \textit{drift-plus-penalty} \cite{Lyapunov}:
	\begin{equation}
	\Delta(t_k) + V \mathbb{E} \bigg[ \sum_{t=t_k}^{t_k+T-1}\nolimits (\bar{\mathcal{P}} - \mathcal{P}(q(t))) \cdot M(t) \bigg| Z(t_k) \bigg], \label{eq:drift-plus-penalty}
	\end{equation}
	where $V$ is an importance weight for quality improvement.
	
	First of all, the upper bound on the drift can be found in the Lyapunov function.
	\begin{align}
	L(t+1) - L(t) &= \frac{1}{2} \Big\{ Z(t+1)^2 - Z(t)^2 \Big\} \nonumber \\
	&= \frac{1}{2} \Big\{ \min\{ Z(t)-M(t) + c, \tilde{Q}-M(t) \}^2  - Z(t)^2 \Big\} \nonumber \\
	&\leq \frac{1}{2} \Big\{ (Z(t)-M(t)+c)^2 - Z(t)^2 \Big\} \label{eq:Lypunov_drift}
	\end{align}
	
	By summing \eqref{eq:Lypunov_drift} over $t=t_k,\cdots,t_k+T-1$, the upper bound in the frame-based Lyapunov function is obtained by
	\begin{equation}
	L(t_k+T) - L(t_k) \leq \sum_{t=t_k}^{t_k+T-1} \bigg\{ Z(t)(c-M(t)) + \frac{1}{2}(c-M(t))^2 \bigg\}.
	\end{equation}
	Thus, according to \eqref{eq:drift-plus-penalty}, minimizing a bound on frame-based \textit{drift-plus-penalty} is equivalent to minimizing
	\begin{align}
	&\mathcal{D}(\alpha(t_k),Q(t_k), {\boldsymbol q}_k, {\boldsymbol M}_k) = \nonumber \\
	&~~~~\mathbb{E}\Bigg[ \sum_{t=t_k}^{t_k+T-1}\nolimits \bigg\{ Z(t)(c-M(t)) + \frac{1}{2}(c-M(t))^2 + V (\bar{\mathcal{P}} - \mathcal{P}(q(t))) \cdot M(t) \bigg\} \bigg| Z(t_k) \Bigg],
	\label{eq:exp_drift-plus-penalty}
	\end{align}
	where $\boldsymbol{q}_k = [q(t_k), q(t_k+1), \cdots, q(t_k+T-1)]$, $\boldsymbol{M}_k = [M(t_k), M(t_k+1), \cdots, M(t_k+T-1)]$ and recall that $Z(t) = \tilde Q - Q(t)$.
	The above minimum is conditioned on $M(t)N_{q(t)}\leq B(\alpha(t_k),t)$ for all $t\in \mathcal{T}_k$.
	This frame-based algorithm is shown to satisfy the queue stability constraint of \eqref{eq:opt_const_queue} while minimizing the objective function of \eqref{eq:opt_metric} in \cite{TAC2013Neely}.
	For any $\alpha(t_k) \in \mathcal{A}(t_k)$, the minimum bound on frame-based drift-plus-penalty can be obtained by
	%For simplicity, we deliberately abuse the notation by
	\begin{equation}
	\mathcal{D}(\alpha(t_k),Q(t_k)) = \min_{{\boldsymbol q}_k, {\boldsymbol M}_k} \mathcal{D}(\alpha(t_k),Q(t_k), {\boldsymbol q}_k, {\boldsymbol M}_k).
	\label{eq:minimum-D-over-q-M}
	\end{equation}
	In Section \ref{sec:qual_chunk_decisions}, we will provide
	an efficient method to find the minimum achieving ${\boldsymbol q}_k$ and ${\boldsymbol M}_k$.
	
	System parameter $V$ in \eqref{eq:exp_drift-plus-penalty} is a weight factor for the term representing the measure of video quality degradation. 
	The value of $V$ is important to control the queue backlogs and quality measures at every time. 
	The appropriate initial value of $V$ needs to be obtained by experiment because it depends on the distribution of caching nodes, the channel environments, the playback rate $c$, and $\tilde{Q}$. 
	Also, $V\geq 0$ should be satisfied.
	If $V<0$, the optimization goal is converted into maximizing the measure of video quality degradation. 
	Moreover, in the case of $V=0$, the user only aims at stacking queue backlogs without consideration of video quality.
	On the other hand, when $V\rightarrow \infty$, users do not consider the queue state, and thus they just pursue to minimize the video quality degradation. 
	$V$ can be regarded as the parameter to control the trade-off between quality and delay, which captures the fact that the user can stack many low-quality chunks or relatively the small number of high-quality chunks in the queue, under the given channel condition.
	
	From \eqref{eq:exp_drift-plus-penalty}, we can anticipate how the algorithm works.
	When the queue is almost empty, i.e. $Z(t) \simeq \tilde{Q}$, the large arrivals $M(t)$ are necessary for the user not to wait the next chunk. 
	In this case, the user prefers the caching node which gives many chunks.
	On the other hand, when the queue backlogs are stacked enough to avoid playback delay, i.e. $Z(t)\simeq 0$, the user would request the high quality level of $\mathcal{P}(q(t))$ without worrying about playback latency. 
	
	With the initial condition of $Q(t_k)$, the user computes $\mathcal{D}(\alpha(t_k),Q(t_k))$ for all $\alpha(t_k) \in \mathcal{A}(t_k)$. 
	Then, the caching node which minimizes $\mathcal{D}(\alpha(t_k),Q(t_k))$ is chosen at the user,
	\begin{equation}
	\alpha^*(t_k) = \underset{\alpha(t_k)\in \mathcal{A}(t_k)}{\argmin} \mathcal{D}(\alpha(t_k),Q(t_k)).
	\label{eq:caching_node_decision}
	\end{equation}
	
	However, the user should estimate the average function value of future queue states $Z(t)$ and decisions of $q(t)$ and $M(t)$ for $t\in \mathcal{T}_k$.
	For finding \eqref{eq:minimum-D-over-q-M}, the frame-based algorithm is formulated based on Markov decision process \cite{TAC2013Neely}, and it can be solved by dynamic programming as following section.
	
	\section{Decisions on Quality Level and Receiving Chunk Amounts}
	\label{sec:qual_chunk_decisions}
	
	The goal of this section is to compute $\mathcal{D}(\alpha(t_k),Q(t_k))$, given the associated caching node $\alpha(t_k)$ and initial queue backlogs $Q(t_k)$.
	
	\subsection{Stochastic shortest path problem}
	
	According to \eqref{eq:exp_drift-plus-penalty}, we can formulate the drift-plus-penalty algorithm of the $k$-th frame as follows:
	\begin{align}
	\{ \boldsymbol{q}_k,\boldsymbol{M}_k \} &= \underset{q,M}{\argmin}~  \mathcal{D}(\alpha(t_k),Q(t_k), {\boldsymbol q}, {\boldsymbol M}) \label{eq:opt2_metric}  \\
	&~~\text{s.t.}~M(t)N_{q(t)} \leq B_k(t) \label{eq:opt2_const1} \\
	&~~~~~~~q(t) \in \mathcal{L}_{l(\alpha(t_k))}, \label{eq:opt2_const2}
	\end{align}
	where $B_k(t)\triangleq B(\alpha(t_k), t)$.
	The problem of \eqref{eq:opt2_metric}-\eqref{eq:opt2_const2} is similar to the stochastic shortest path problem based on Markov decision process.
	In the network model, $B_k(t)$ and $Z(t)$ (i.e., $Q(t)$) are given before making decisions of $q_k(t)$ and $M_k(t)$ at every time $t$.
	
	The queue backlog $Z(t)$ represents the current state which satisfies the Markov property.
	Define $\mathcal{Z}=\{0,1,\cdots,\tilde{Q} \}$ as the state space of the user queue.
	It is reasonable to set $\tilde{Q}$ be the arbitrarily predefined maximum queue backlog
	because the queue size is finite in practical system.
	The action set is defined as $\Theta(t) = \{M(t),q(t) \}$.
	Then, incurred cost at $t \in \mathcal{T}_k$ can be formulated by
	\begin{align}
	g_k(Z(t),\Theta(t)) &= Z(t) (c-M(t)) + \frac{1}{2} (c-M(t))^2 + V(\bar{\mathcal{P}} - \mathcal{P}(q(t)))M(t).
	\end{align}
	
	The transition probabilities from $Z(t)$ to $Z(t+1)$ can be defined for all states $i$ and $j$ as
	\begin{equation}
	P_{ij}(\Theta, b) = {P}\{ Z(t+1)=j | Z(t) = i, \Theta(t) = \Theta, B_k(t) = b \}.
	\end{equation}
	Since the next state $Z(t+1)$ is deterministic given $Z(t)$ and action $M(t)$, it can be seen that $P_{ij}(\Theta,b) \in \{0,1\}$.
	
	\subsection{Probability mass function of $B_k(t)$}
	
	The constraint \eqref{eq:opt2_const1} indicates that the maximum number of chunks which the user can receive depends on the random network event $B_k(t)$ and the decision of quality $q(t)$.
	It notifies that decisions on $q(t)$ and $M(t)$ should jointly made as well as the probability distribution of the random network event $B_k(t)$ is required.
	
	Define a random variable $Y=\log_2(1+aX)$, where $X$ is a chi-square random variable and $a$ is a constant.
	Then, we can obtain $P\{Y \geq y\}$, as given by
	\begin{equation}
	P\{Y \geq y\} = \exp\Big\{ -\frac{1}{2a}(2^y-1) \Big\}.
	\end{equation}
	Since $|h(\alpha,t)|^2$ follows the chi-squared distribution and a random variable $B_k(t)$ is a nonnegative integer, the probability mass function of $B_k(t)$ is found as follows:
	\begin{align}
	P\{B_k(t)=b\} &=P\{t_c R(\alpha(t_k), t) \geq b \} - P\{t_c R(\alpha(t_k), t) \geq b+1\} \\
	&=e^{1/2\Gamma} D(\alpha, t) \Bigg\{ \exp \bigg\{ -\frac{2^{b/t_c\mathcal{W}}}{2\Gamma D(\alpha, t)} \bigg\} - \exp \bigg\{ -\frac{2^{(b+1)/t_c\mathcal{W}}}{2\Gamma D(\alpha, t)} \bigg\} \Bigg\}.
	\end{align}
	
	\subsection{Dynamic programming}
	\label{subsec:dynamic_program}
	
	Given $\alpha(t_k)$ and $Z(t_k)$, the user observes the queue state $Z(\tau)$ and the random network event $B_k(\tau)$, and decides the action $\Theta(\tau)$ for each time slot $\tau \in \mathcal{T}_k$.
	Then, the minimum incurred cost based on measurements of $B_k(\tau)$ and $Z(\tau)$ is %given by
	\begin{equation}
	G_k(\tau,z_0,b_0) = \underset{\Theta}{\min}~\mathbb{E} \left[ \sum_{t=\tau}^{t_k+T-1} g_k(Z(t),\Theta(t)) \Big| Z(\tau)=z_0, B_k(\tau)=b_0 \right],
	\end{equation}
	conditioned on $M(\tau) N_{q(\tau)} \leq B_k(\tau)$.
	
	Let $J_k(\tau,z_0)$ be the marginalized function of $G_k(\tau,z_0,b_0)$ over all possible $B_k(\tau)=b_0$, and it can be approximated into
	\begin{equation}
	J_k(\tau,z_0) = \sum_{b_0=0}^{B_{\text{max}}} P\{ B_k(\tau)=b_0 \} G_k(\tau,z_0,b_0),
	\end{equation}
	where $B_{\text{max}}$ is a nonnegative integer such that $P\{B \geq B_{\text{max}} \} \approx 0$.
	The dynamic programming provides the action that minimizes the following cost as given by \cite{DynamicProgram}
	\begin{align}
	G_k(\tau,z_0,b_0) &= \underset{ \Theta }{\min}~\mathbb{E} \Big[ g_{k}(Z(\tau)=z_0,\Theta(\tau)) + \sum_{y\in \mathcal{Z}} P_{z_0,y}(\Theta(\tau),b_0) \cdot G_k(\tau+1, y,B_k(\tau+1)) \Big] \label{eq:DP1} \\
	&= \underset{\Theta}{\min} \Big[ g_{k}(Z(\tau)=z_0,\Theta(\tau)) + J_k(\tau+1,Z(\tau+1)) \Big], \label{eq:DP2}
	\end{align}
	where the expectation of \eqref{eq:DP1} is with respect to $\{ B_k(t):~\tau+1 \leq t \leq T-1 \}$ and $Z(\tau+1) = \min \{z_0 + c,\tilde{Q} \} - M(\tau)$.
	The minimum cost is obtained over all $\Theta(\tau)$ such that $M(\tau) N_{q(\tau)} \leq B_k(\tau)=b_0$.
	
	Given $B_k(t)=b_0$, the user can find the minimum value of \eqref{eq:DP2} by greedily testing all joint combinations of decisions on $q(t)$ and $M(t)$.
	For example, let there exists $L=2$ quality levels and $q\in \{1,2\}$ 
	correspond to the file size of $N\in \{10,20\}$
	If $B_k(t) \in [20$:$30)$ Kbits, where $[a$:$b) \triangleq \{a, a+1, \cdots, b-1\}$ for simplicity, then there are four possible decisions: 1) $M(t)=0$, 2) $q(t)=1,M(t)=1$, 3) $q(t)=1,M(t)=2$, and 4) $q(t)=2,M(t)=1$. 
	The user computes costs for all those possible decision cases and picks the minimum one as an optimal cost.
	
	We set the end time slot of the $k$-th frame as $t_{k+1}=t_k+T$, which is the start time of the $k+1$-th frame.
	To find the optimal costs $\mathbf{J}_k(t) = [J_k(t,0),\cdots,J_k(t,\tilde{Q})]$ for $t\in \mathcal{T}_k$ by using dynamic programming equation \eqref{eq:DP2}, the end costs of $\mathbf{J}_k(t_{k+1})$ are required.
	Since the playback delay occurs at the end state when the accumulated chunk amounts are smaller than the departure quantity, i.e., $Q(t_{k+1})<c$ and $Z(t_{k+1}) > \tilde{Q}-c$.
	Therefore, the end costs for those states, i.e., $J_k(t_{k+1},z)$ for $z\in\{ \tilde{Q}-c+1, \cdots, \tilde{Q} \}$ should be very large.
	Even when $Q(t_{k+1}) \geq c$, the more chunks are accumulated, the more likely there will be no playback delays. 
	In this sense, $J_k(t_{k+1},i) \geq J_k(t_{k+1},j)$ for $i \geq j$ is preferred for all $i,j \in \{0,\cdots,\tilde{Q}-c \}$.
	Especially, as a large number of chunks are received in the queue, the effect of additional chunks to avert queue emptiness would be significantly decreased.
	Therefore, $J_k(t_{k+1},i)$ for $i=\{1,\cdots \tilde{Q} \}$ are arbitrarily modeled as the truncated form of exponential distribution. 
	Thus, we can set the end costs for all states as follows:
	\begin{align}
	&J_k(t_{k+1},z) = A,~\forall z \in \{\tilde{Q}-c+1, \cdots, \tilde{Q} \} \\
	&J_k(t_{k+1},i) = 10^{-3} \cdot A  \mu e^{-\mu \cdot (\tilde{Q}-i)},~\forall i \in \{0,\cdots,\tilde{Q}-c \},
	\end{align}
	where $A$ is a predefined large constant to give penalties for playback delay occurrences and $\mu$ is the exponential distribution coefficient.
	
	Given the end costs $\mathbf{J}_k(t_{k+1})$, the optimal costs $\mathbf{J}_k(t)$ for all $t \in \mathcal{T}_k$ can be obtained by backtracking the shortest path based on the dynamic programming equation \eqref{eq:DP2}.
	Then, when the queue backlog at time $t=t_k$ is $Z(t_k)$, $J_k(t_k,Z(t_k))$ becomes the averaged drift-plus-penalty term \eqref{eq:exp_drift-plus-penalty}, i.e., $\mathcal{D}_k$.
	Then, after the user finds all the averaged drift-plus-penalty terms for all $\alpha \in \mathcal{A}(t)$ by using dynamic programming, the user determines the caching node to receive the desired video file by comparing all drift-plus-penalty terms, as described in \eqref{eq:caching_node_decision}.
	
	\subsection{Decisions of quality and chunk amounts}
	\label{subsec:quality_chunk_decisions}
	
	After determining the caching node $\alpha(t_k)$, the user should choose the video quality and the number of chunks to receive for every time slot $t\in \mathcal{T}_k$, depending on time-varying channel conditions and its queue state.
	For this goal, we can simply use the principle of optimality in the dynamic programming algorithm \cite{DynamicProgram}, which argues that if the optimal policy $\Theta^* = \{\Theta^*(t_k), \cdots, \Theta^*(t_k+T-1) \}$ 
	is a solution of the stochastic shortest path problem, then the truncated policy $\{\Theta^*(t_k+j), \cdots, \Theta^*(t_k+T-1) \}$ is optimal for the subproblem over $t\in \{t_k+j,\cdots, t_k+T-1 \}$, where $0\leq j \leq T-1$.
	
	Based on this principle of optimality, the user can make the optimal decisions of $q^*(t)$ and $M^*(t)$ for $t\in \mathcal
	{T}_k$ by using the minimum costs obtained while performing dynamic programming for caching node decision $\alpha^*(t_k)$.
	When deciding $q^*(t)$ and $M^*(t)$, the channel gain can be observed, e.g. $B(\alpha(t_k),t) = b_0$,
	so the optimal action $\Theta^*(t)$ is deterministic given $Z(t)=z_0$ and $B_k(t) = b_0$ which provides the minimum cost $G_k(t,z_0,b_0)$, as given by
	\begin{equation}
	\Theta_k^*(t,z_0,b_0) = \underset{ \Theta }{\argmin}~ \Big[ g_{k}(Z(t)=z_0,\Theta) + J_k(t+1,Z(t+1)) \Big],
	\end{equation}
	conditioned on $M(t)N_{q(t)} \leq b_0$ for $t \in \mathcal{T}_k$.
	
	Thus, the user should store the optimal actions $\Theta_k^*(t,z,b)$ for all $t\in \mathcal{T}_k$, $z\in \mathcal{Z}$ and $b \in \mathcal{B} = \{0,1,\cdots,B_{\text{max}} \}$ to deal with all possible random network events.
	Simply, $T\cdot \tilde{Q} \cdot (B_{\text{max}}+1)$ actions are required, but some of channel realizations can give the same optimal action.
	Again, consider the example of $L=2$ quality levels and $q\in \{1,2\}$ corresponding to the file size of $N\in \{10,20\}$ in Kbits.
	Then, any $B=b \in [20$:$30)$ Kbits allows four combinations of decisions of $q(t)$ and $M(t)$, as explained in Section \ref{subsec:dynamic_program}, and the user is enough to store the only one optimal action for all $B=b \in [20$:$30)$ Kbits.
	In this sense, define $\mathcal{N}_B$ subsets of $\mathcal{B}$ denoted by $\mathcal{B}_n$ for $n \in \{1,\cdots, \mathcal{N}_B \}$, as follows:
	\begin{align}
	&\bigcup_{n=1}^{\mathcal{N}_B} \mathcal{B}_n = \mathcal{B} \\
	&\mathcal{B}_n \cap \mathcal{B}_m = \phi,~\forall n\neq m,~n,m\in \{ 1, \cdots, \mathcal{N}_B \} \\
	&\Theta^*_k(t,z,b_1) = \Theta^*_k(t,z,b_2),~\forall b_1, b_2 \in \mathcal{B}_n.
	\end{align}
	Thus, the user needs to store $T \cdot \tilde{Q} \cdot \mathcal{N}_B$ actions.
	The whole steps for video delivery decisions on caching node, video quality, and receiving chunk amounts are presented in Algorithm \ref{alg:video_delivery}.
	
	\begin{algorithm}[t!]
		\caption{Dynamic video delivery decisions on $\boldsymbol{\alpha}$, $\boldsymbol{q}$, and $\boldsymbol{M}$ in different timescales
			\label{alg:video_delivery}}
		\begin{algorithmic}[1]
			\Require{\\
				\begin{itemize}
					\item $V$: parameter for streaming quality-delay trade-offs
					\item $\tilde{Q}$: threshold for queue backlog size
					\item $K$: the number of caching node decisions
					\item $T$: the time interval of updating caching node decision
				\end{itemize}
			}
			\State{$t=0$ // $KT-1$: number of discrete-time operations}
			
			\While{$k \leq K$}{
				\State{$t_k = (k-1)T$: time for the $k$-th caching node decision}
				\State{Observe $Z(t_k)$ and find $\mathcal{A}(t_k)$}
				\State{Compute $\mathcal{D}_k(\alpha(t_k),Q(t_k))$ by using dynamic programming equation \eqref{eq:DP2} and store $\Theta_k^*(t,z,b)$ for every $\alpha(t_k) \in \mathcal{A}(t_k)$, $z\in \mathcal{Z}$ and $b\in \mathcal{B}$}.
				\State{Make a decision of $\alpha^*(t_k)$ by using \eqref{eq:caching_node_decision}}
				\For{$t=t_k:t_k+T-1$}{
					\State{Observe $Z(t)$ and $B_k(t)$}
					\State{Make a decision of $\Theta_k^*(t,Z(t),B_k(t))$}
					\EndFor}
				%\EndFor
			}
			\EndWhile
		\end{algorithmic}
	\end{algorithm}
	
	\subsection{Computational complexity of dynamic programming}
	
	To determine the optimal policy at each time slot, it seems that at least $\tilde{Q} B_{\text{max}}$ computations are required, but some of channel realizations can perform the same computation as seen in Section \ref{subsec:quality_chunk_decisions}. 
	Since all realizations $b\in \mathcal{B}_n$ not only give the same $\Theta_k^*(t,Z(t),b)$ but also make the same computations of $g_{k}(Z(t)=z_0,\Theta(t)) + J_k(t+1,Z(t+1))$ for all possible combinations of $q(t)$ and $M(t)$, $\tilde{Q} \mathcal{N}_B$ computations are required at least.
	
	However, in most of random network events, more computations are required to take the minimum function in \eqref{eq:DP2}.
	As shown in the example of $L=2$ quality levels and 
	$q\in \{1,2\}$ corresponding to the file size of 
	$N\in\{10,20\}$ in Kbits, there are four combinations of decisions of $q(t)$ and $M(t)$ when 
	$B=b \in [20$:$30)$ Kbits.
	Let the average number of these decision combinations of $q(t)$ and $M(t)$ for all $B_k(t) \in \mathcal{B}$ be $\mathcal{N}_{\theta}$, then total $\tilde{Q} \mathcal{N}_B \mathcal{N}_{\theta}$ computations are required at each time slot in dynamic programming.
	
	Here, $\mathcal{N}_B$ and $\mathcal{N}_{\theta}$ obviously depend on $B_{\text{max}}$, $L$ and $N_q$ for $q \in \mathcal{L}_L$.
	There are not many versions of the identical video of different quality levels, i.e. $L$ is small in general, and $N_q$ is not controllable unless the video encoding scheme is changed.
	On the other hand, $B_{\text{max}}$ increases as transmit SNR grows, therefore large SNR could result in huge computational complexity as well as large number of registers to store the optimal costs for decisions of quality and chunk amounts.
	However, the streaming user can receive a large number of high-quality chunks enough to avoid queue emptiness in the sufficiently large transmit SNR region.
	Considering that the proposed video delivery scheme targets the streaming user who is worrying about playback delays as well as video quality degradation, however, huge complexity burden for large transmit SNR is out of scope in targeting scenarios.
	Thus, $\mathcal{N}_B$ and $\mathcal{N}_{\theta}$ are expected not much large in our targeting scenarios where adjustments of the tradeoff between playback delay and video quality are necessary, so computational complexity for dynamic programming can be somewhat limited.
	
	\begin{table}[t!]
		\small
		%\footnotesize
		\caption{System Parameters}
		\label{table:parameters}
		\begin{center}
			\scalebox{1}{
				\begin{tabular}{l|c}
					\toprule
					Description & Value \\
					\midrule [1.0pt]
					No. of quality levels ($L$) & 3 \\
					Default PPP intensity ($\lambda$) & 0.4 \\
					Time interval of caching node decisions ($T$) & 5 \\ 
					User radius ($R$) & 50 m \\ 
					Caching probabilities ($\mathbf{p}=[p_1,\cdots,p_L]$) & $[\frac{4}{7}, \frac{2}{7}, \frac{1}{7}]$ \\
					Transmit SNR ($\Psi$) & 20 dB \\ 
					INR ($\Upsilon$) & 5dB \\
					Minimum probability of finding the caching node ($\eta_{\text{min}}$) & 0.99 \\
					Queue departure ($c$) & 1 \\
					Bandwidth ($\mathcal{W}$) & 1 MHz \\
					Coherence time ($t_c$) & 5 ms \\
					End cost coefficient ($A$) & $10^4$ \\
					End cost coefficient ($\mu$) & 1 \\
					$\tilde{Q}$ & 100 \\
					$B_{\text{max}}$ & 52 kbits \\ 
					$V$ & 0.015 \\
					\bottomrule
				\end{tabular}
			}
		\end{center}
	\end{table}

	\section{Simulation Results}
	\label{sec:simulation}
	
	In this section, we show that the proposed algorithm for dynamic video delivery policy works well with video files of different quality levels in wireless caching network.
	Simulation parameters are listed in Table \ref{table:parameters}, and these are used unless otherwise noted.
	The proposed technique can be applied to any distribution model for caching nodes, but we suppose that caching nodes which store the desired content are modeled as an independent PPP with an intensity of $\lambda$, which is generally assumed for researches of wireless caching networks \cite{caching:ICC2015Blaszczyszyn, CL2017Chen, caching:TWC2016Chae}.
	Then, the PPP intensity of type-$l$ caching nodes becomes $\lambda p_l$, where $p_l$ denote the caching probability of the video which can be encoded into any quality in $\mathcal{L}_l$.
	Therefore, larger $p_l$, more caching nodes of type-$l$ around the streaming user.
	Based on the network model described in Fig. \ref{fig:network_model}, the user is slowly moving towards certain direction.
	In practice, the channel condition between the user and the caching node delivering the desired video could be varying due to Doppler shift as the user is moving, but this effect is not captured in this paper. 
	Peak-signal-to-noise ratio (PSNR) is considered as a video quality measure, and quality measures and file sizes depending on quality levels are supposed as $\mathcal{P}(q) = [34,~36.64,~39.11]$ dB and $N(q) = [2621,~5073,~10658]$ Kbits which are obtained from real-world video traces~\cite{ICTC2015Kim}.
	Since we assume that $\eta_{\text{min}} = 0.99$ and $t_0 \mathcal{W} \log_2(1+\gamma_{\text{min}}) = N(1)$, the minimum intensity of PPP distributions to satisfy the performance criterion of $\{\gamma_{\text{min}}, \eta_{\text{min}} \}$ should be $\lambda_{\text{min}} = 0.1113$. 
	
	To verify the advantages of the proposed algorithm, this paper compares the proposed one with three other schemes:
	\begin{itemize}
		\item `Strongest': The user receives the desired video file from the caching node whose channel condition is the strongest among $\mathcal{A}(t_k)$ at time slots of $t_k$, for $k\in\{1,2,\cdots \}$.
		Decisions of $q(t)$ and $M(t)$ are made based on dynamic programming results.
		
		\item `Highest-Quality': The user receives the desired video file from the caching node which can provide the highest-quality file among $\mathcal{A}(t_k)$ at time slots of $t_k$, for $k\in \{1,2,\cdots \}$.
		Decisions of $q(t)$ and $M(t)$ are made based on dynamic programming results.
		
		\item `One-Step': The user decides the caching node for video delivery based on the frame-based Lyapunov optimization theory. 
		However, decisions of $q(t)$ and $M(t)$ are made by minimizing the incurred cost only at each slot $t$ without using dynamic programming results.
	\end{itemize}
	In summary, performance comparisons with `Strongest' and `Highest-Quality' can show the effects of caching node decision based on Lyapunov optimization, and comparison with `One-Step' can specify the advantage of using Markov decision process and dynamic programming for decisions on video quality and the amounts of receiving chunks.
	
	\begin{figure}[t]
		\minipage{0.47\textwidth}
		\includegraphics[width=\linewidth]{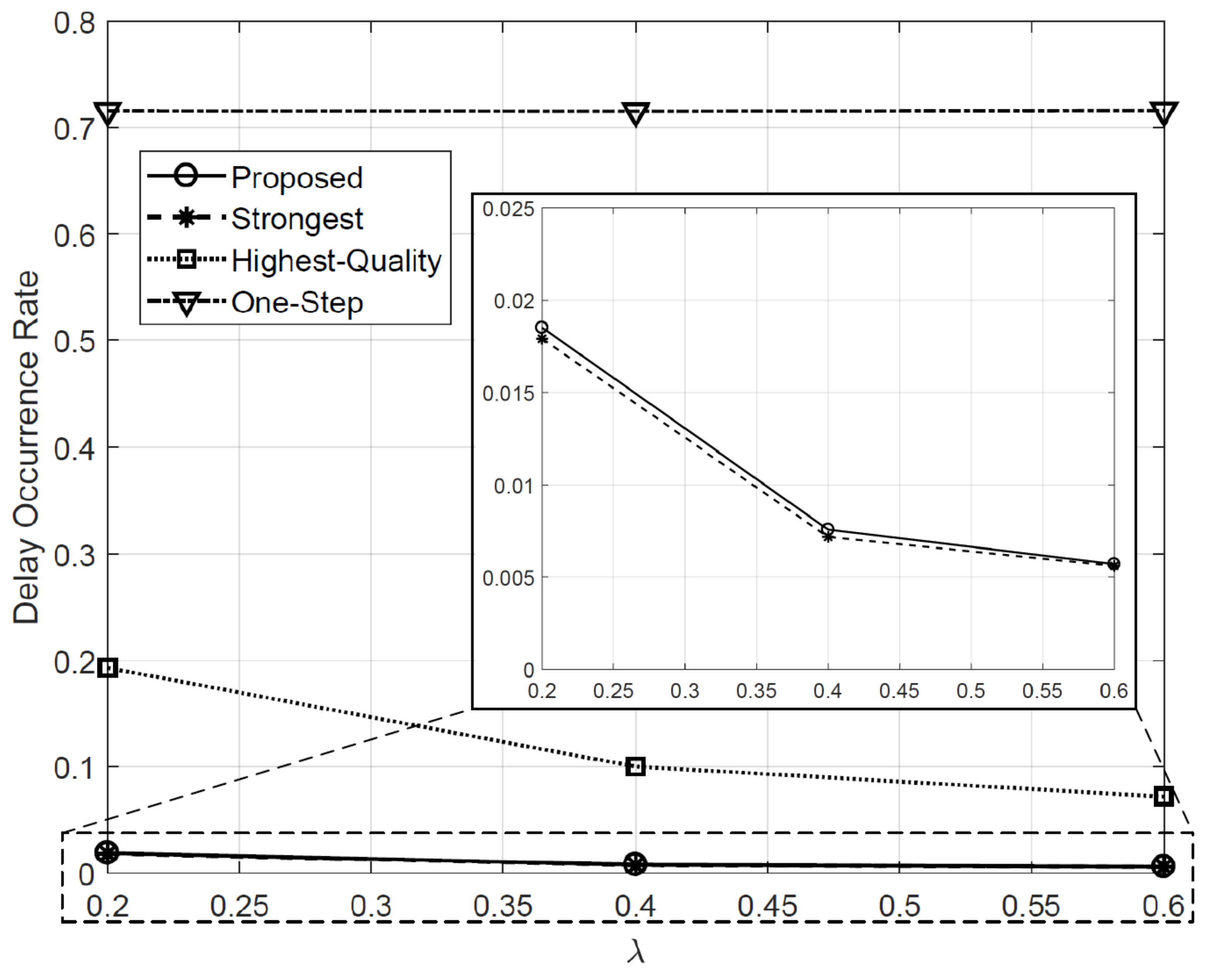}
		\caption{Delay occurrence rates over $\lambda$}
		\label{fig:delay_lambda}
		\endminipage\hfill
		\minipage{0.47 \textwidth}
		\includegraphics[width=\linewidth]{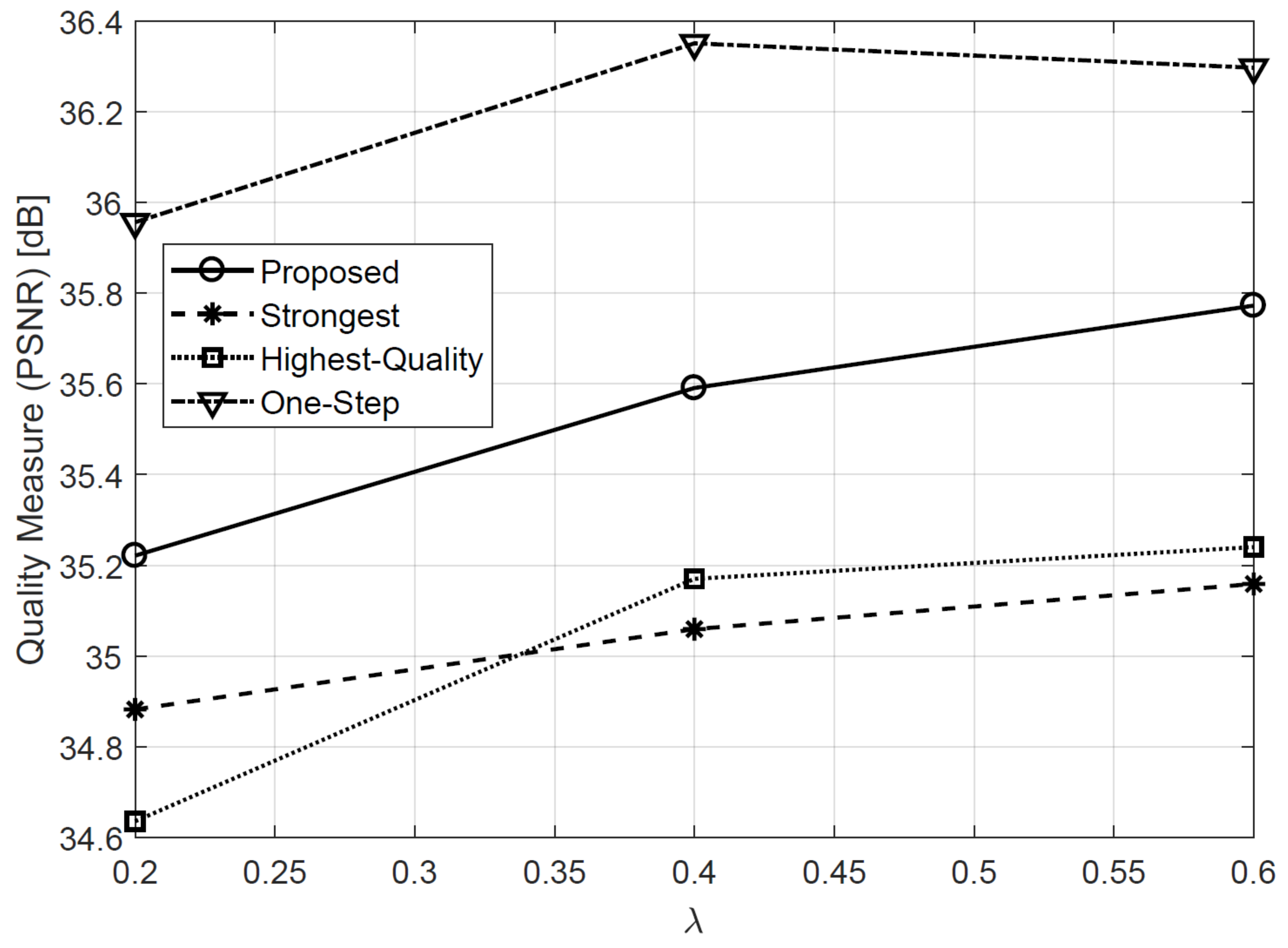}
		\caption{Video quality measures over $\lambda$}
		\label{fig:quality_lambda}
		\endminipage
	\end{figure}
	
	%\begin{figure} [t!]
	%	\centering
	%	\includegraphics[width=0.6\columnwidth]{delay_lambda.pdf}
	%	\caption{Delay occurrence rates over $\lambda$}
	%	\label{fig:delay_lambda}
	%\end{figure}
	
	%\begin{figure} [t!]
	%	\centering
	%	\includegraphics[width=0.6\columnwidth]{quality_lambda.pdf}
	%	\caption{Video quality measures over $\lambda$}
	%	\label{fig:quality_lambda}
	%\end{figure}
	
	\subsection{Caching node distribution}
	\label{subsec:simul-node}
	
	At first, impacts of the PPP intensity, i.e. how many caching nodes are distributed around the streaming user, are shown in Figs. \ref{fig:delay_lambda} and \ref{fig:quality_lambda}, which give the plots of playback delay occurrence rates and average video quality measures per received chunk versus $\lambda$, respectively. 
	`Strongest' is likely to receive many chunks from the caching node whose channel condition is the strongest, so this scheme accumulates queue backlogs enough to avoid playback delays.
	Therefore, `Strongest' shows the best delay performance but its gain over the proposed one is very small, as shown in enlarged plots in Fig. \ref{fig:delay_lambda}.
	There are two reasons. 
	The first one is that even though the channel condition of certain caching node at $t_k$ is the strongest, after that it could not be the strongest due to time-varying channels and user mobility.
	Second, the delay performance does not increase in proportional to the number of chunks accumulated in the queue.	
	If enough chunks are already in the queue to prevent playback delays, then the delay performance is not dramatically improved as additional chunks arrive.
	Similar delay occurrence rates of the proposed technique and `Strongest' in Fig. \ref{fig:delay_lambda} show that the proposed scheme can accumulate chunks in the queue enough to avert playback delays.
	
	On the other hand, since `Highest-Quality' pursues the video quality when choosing the caching node for video delivery, it gives better quality performance than `Strongest' with large $\lambda$.
	However, when $\lambda$ is small, the caching node chosen by `Highest-Quality' is likely to be much distanced from the streaming user and its channel condition would be usually too bad to deliver the high-quality video.
	Therefore, even though the caching node chosen by `Highest-Quality' could provide the high quality level, the user requests large number of low-quality chunks owing to less accumulated backlogs.
	Since we assume that the user cannot achieve any quality-of-service when delay occurs, the quality performance of `Highest-Quality' is even worse than that of `Strongest' with small $\lambda$.
	As the proposed technique determines to associate with the caching node by balancing the video quality and channel condition, the proposed one can provide better quality than both `Strongest' and `Highest-Quality', as shown in Fig. \ref{fig:quality_lambda}. 
	`One-Step' gives %the best quality performance 
	the highest average quality measure per received chunk but it suffers from much more frequent delay occurrences compared to other schemes.
	Considering that streaming users are much more sensitive to playback delays, `One-Step' is not appropriate for practical systems.
	From the result of `One-Step', we can see that the merit of using dynamic programming which stochastically reflects future subsequent decisions is very large when determining the video quality and the number of receiving chunks.
	In addition, even when $\lambda = 0.6$, PPP intensity of highest-quality videos ($q=3$) becomes $\lambda p_3 = 0.0857 < \lambda_{\text{min}}=0.1113$. 
	Therefore, the highest-quality level is rarely selected and the average quality measures of all schemes are much lower than the highest-quality measure ($\mathcal{P}(q=3)=39.11\text{dB}$).
	
	\begin{figure}[t]
		\minipage{0.47\textwidth}
		\includegraphics[width=\linewidth]{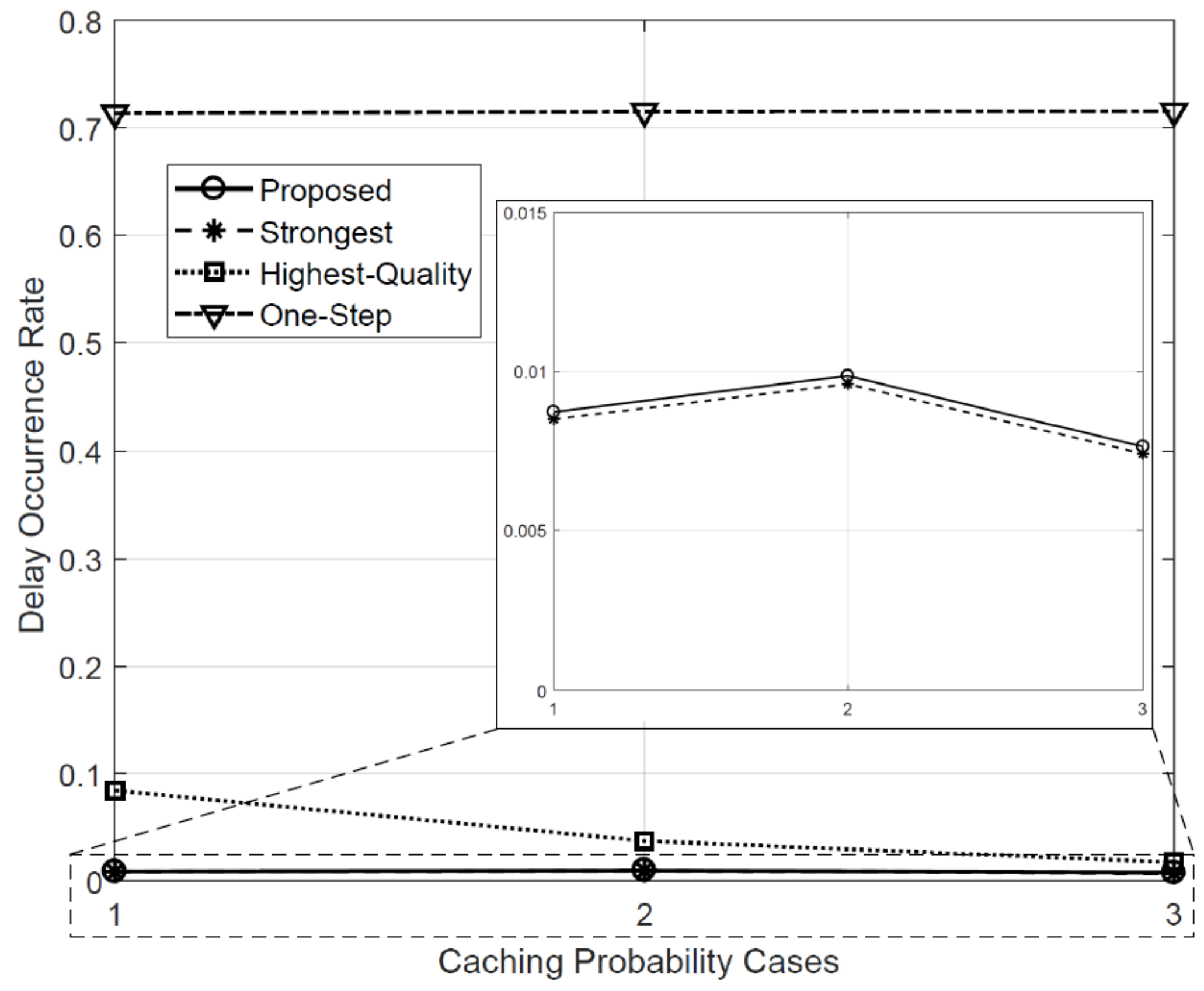}		\caption{Delay occurrence rates over caching policies}
		\label{fig:delay_prob}
		\endminipage\hfill
		\minipage{0.47\textwidth}
		\includegraphics[width=\linewidth]{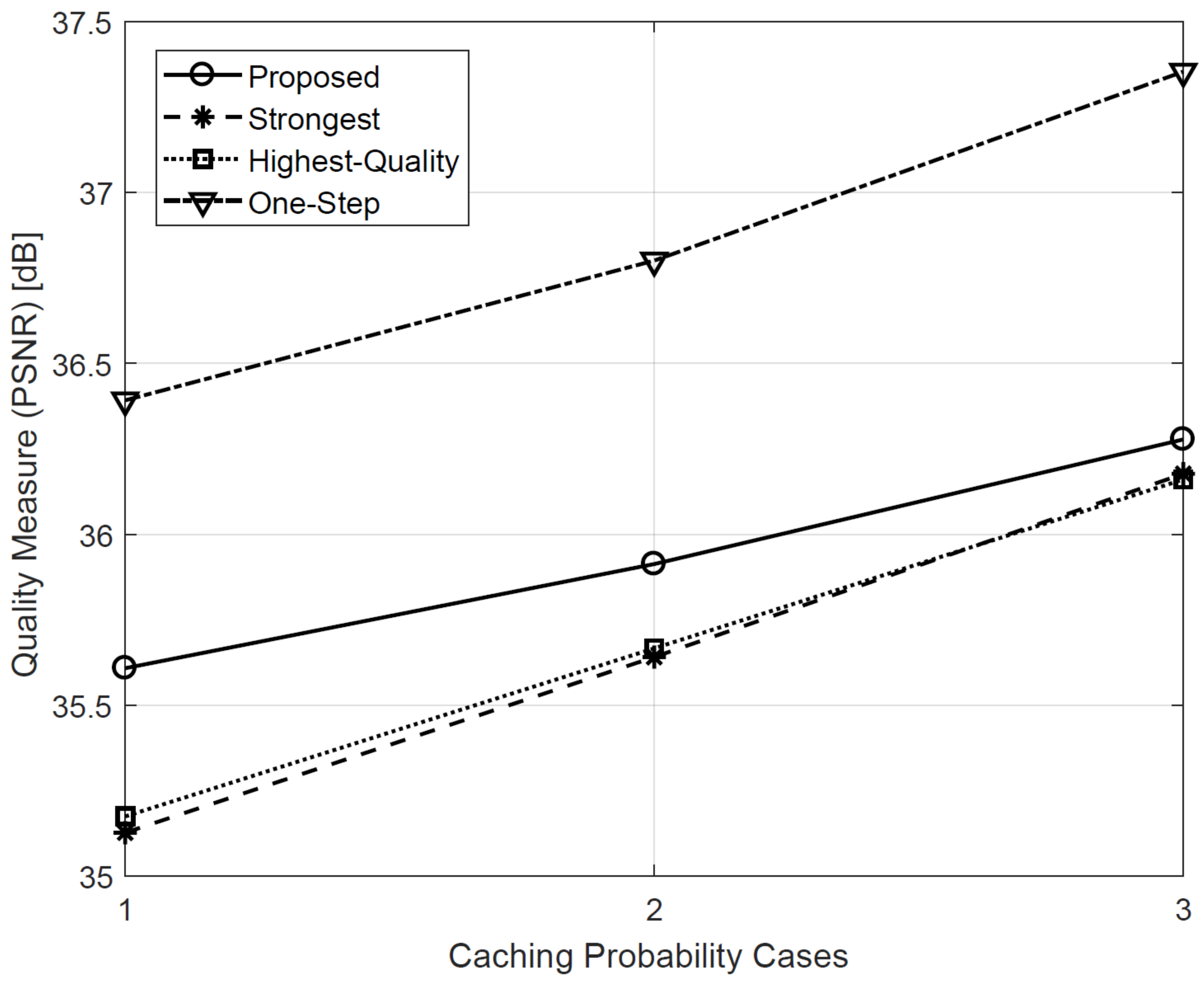}
		\caption{Video quality measures over caching policies}
		\label{fig:quality_prob}
		\endminipage
	\end{figure}
	
	%\begin{figure} [t!]
	%	\centering
	%	\includegraphics[width=0.6\columnwidth]{delay_prob.pdf}
	%	\caption{Delay occurrence rates over $\lambda$}
	%	\label{fig:delay_prob}
	%\end{figure}
	
	%\begin{figure} [t!]
	%	\centering
	%	\includegraphics[width=0.6\columnwidth]{quality_prob.pdf}
	%	\caption{Video quality measures over $\lambda$}
	%	\label{fig:quality_prob}
	%\end{figure}
	
	\subsection{Uniform and nonuniform caching probabilities}
	\label{subsec:simul-cachingProb}
	
	We set three cases of caching probabilities for the video file with different quality levels, as follows:
	\begin{itemize}
		\item Case 1: $p_1=4/7,~p_2=2/7,~p_3=1/7$
		\item Case 2: $p_1=1/3,~p_2=1/3,~p_3=1/3$
		\item Case 3: $p_1=1/7,~p_2=2/7,~p_3=4/7$
	\end{itemize}
	Note that Case 2 corresponds to the uniform caching probability case and Case 1 and Case 3 are nonuniform.
	In Case 3, the streaming user is more likely to receive high-quality video than other cases, on the other hand, Case 1 represents an environment where there are not many caching nodes which can provide the high-quality video around the user.
	The performances of playback delay and quality measure depending on those cases of caching probabilities are shown in Figs. \ref{fig:delay_prob} and \ref{fig:quality_prob}, respectively.
	
	In Fig. \ref{fig:delay_prob}, delay incidence of `Highest-Quality' definitely decreases as $p_1$ decreases and $p_3$ grows, because the caching nodes storing the high-quality video are likely to be near to the streaming user.
	However, since the distribution density of all caching nodes does not change according to the probabilistic caching policy which satisfies $p_1+p_2+p_3=1$, the delay performance of `Strongest' is not influenced much by different caching policies.
	For the `Strongest' scheme, any caching probability case can deliver as many low-quality chunks of small size as possible when there are too few chunks in queue so the playback delay is about to occur.
	In this sense, the proposed technique shows almost the same delay performance as `Strongest', because the proposed one strongly limits the playback delay compared to quality improvement.
	
	The average quality measures of all schemes increase as $p_1$ decreases and $p_3$ grows as shown in Fig. \ref{fig:quality_prob}.
	Even though `Highest-Quality' pursues the video quality, its average quality measure per received chunk does not differ much from that of `Strongest' for any caching probability case owing to its poor delay performance.
	As we have seen in Section \ref{subsec:simul-node}, queue backlogs do not accumulate much in the `Highest-Quality' scheme, therefore the user usually requests the small number of low-quality chunks.
	Especially in Case 3, caching nodes storing the highest-quality video are distributed more than nodes of other types, therefore the caching node whose channel condition is the strongest among candidate nodes would be highly probable to be type 3.
	Thus, the difference between quality performances of `Strongest' and `Highest-Quality' is not large.
	%Compared to `Strongest' and `Highest-Quality', the proposed scheme can provide higher quality even with the very low delay occurrence rate.
	
	%However, delay performances of the proposed scheme and `Strongest' are not largely influenced by caching probability cases, because any case can deliver as many low-quality chunks of small size as possible when there are too few chunks in queue so the playback delay is about to occur.
	%On the other hand, since `Highest-Quality' pursues the video quality rather than averting playback latency especially in Case 3, this scheme chooses the caching node storing the high-quality file, even though its channel condition is not good.
	%Thus, delay performance of `Highest-Quality' is largely degraded in Case 3.
	
	The performance rankings in Figs. \ref{fig:delay_prob} and \ref{fig:quality_prob} among comparison techniques are consistent with the results of Figs. \ref{fig:delay_lambda} and \ref{fig:quality_lambda}. 
	Compared to those comparison schemes, the proposed technique provides quite high average video quality, while limiting delay occurrence rate as low as `Strongest'.
	Thus, the proposed scheme can be said to smooth out the tradeoff between quality and playback delay and to achieve both goals.
	As observed here, `One-Step' provides higher quality than the proposed one but its delay performance is too poor to achieve user satisfaction.
	
	\begin{figure}[t]
		\minipage{0.47\textwidth}
		\includegraphics[width=\linewidth]{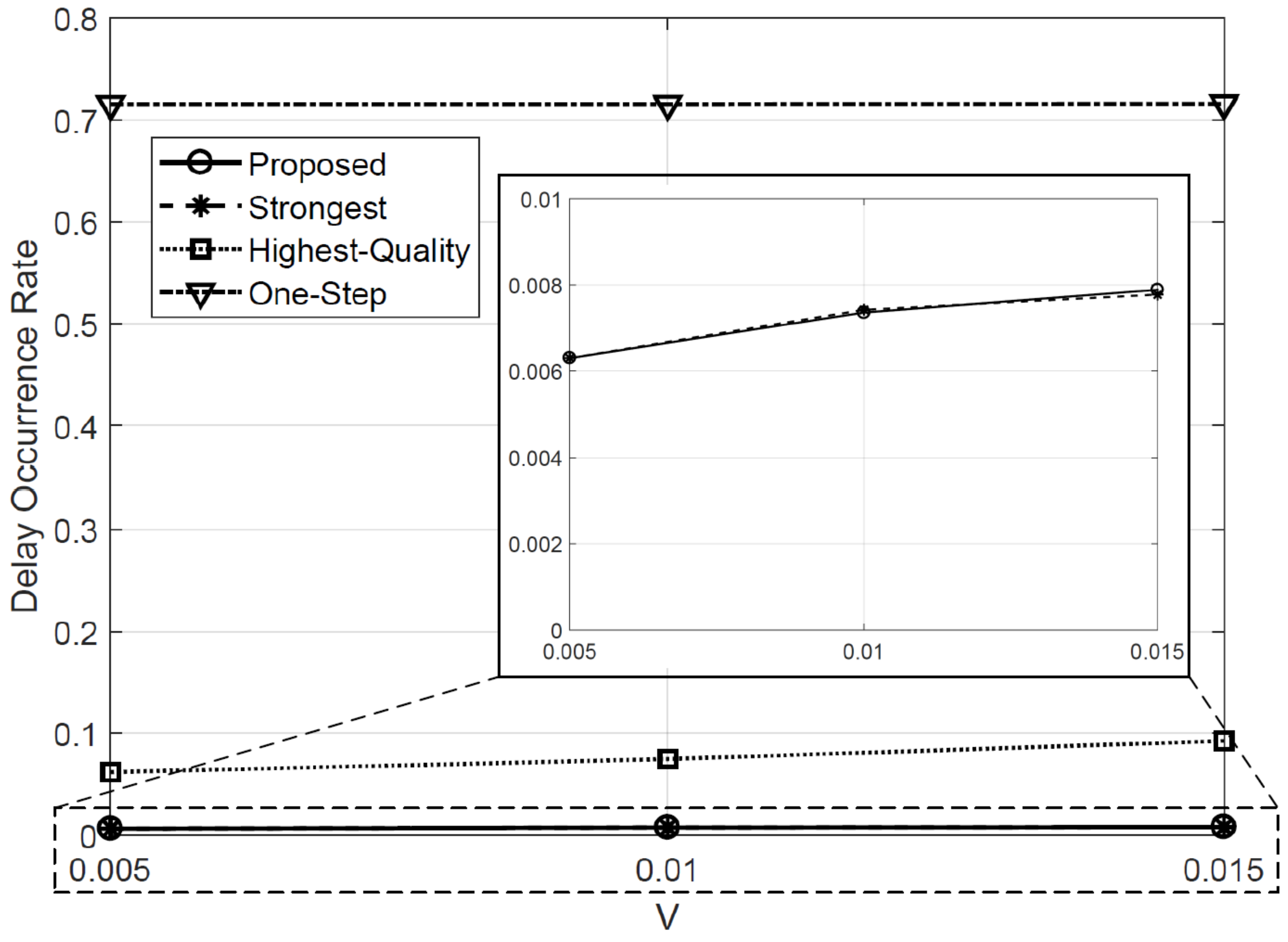}
		\caption{Delay occurrence rates over $V$}
		\label{fig:delay_V}
		\endminipage\hfill
		\minipage{0.47\textwidth}
		\includegraphics[width=\linewidth]{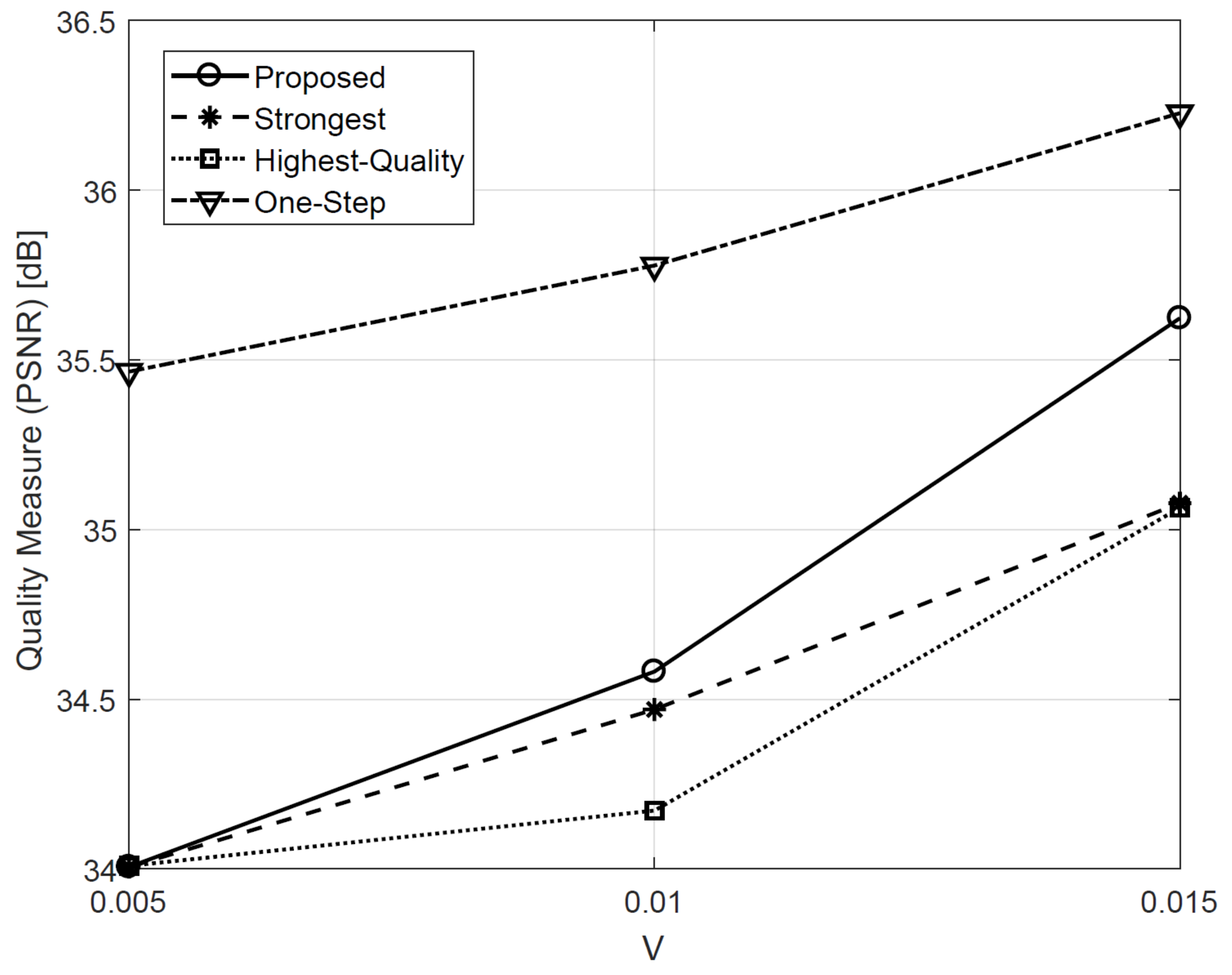}
		\caption{Video quality measures over $V$}
		\label{fig:quality_V}
		\endminipage
	\end{figure}
	
	%\begin{figure} [t!]
	%	\centering
	%	\includegraphics[width=0.6\columnwidth]{delay_V.pdf}
	%	\caption{Delay occurrence rates over $V$}
	%	\label{fig:delay_V}
	%\end{figure}
	
	%\begin{figure} [t!]
	%	\centering
	%	\includegraphics[width=0.6\columnwidth]{quality_V.pdf}
	%	\caption{Video quality measures over $V$}
	%	\label{fig:quality_V}
	%\end{figure}
	
	\subsection{System parameter $V$}
	\label{subsec:simul-V}
	
	Since $V$ has a role to weigh quality maximization compared to averting playback delay in Lyapunov optimization problem, delay occurrence rates increase and the expected quality measures of all techniques become improved, as $V$ grows, as shown in Figs. \ref{fig:delay_V} and \ref{fig:quality_V}, respectively.
	Therefore, we can control the tradeoff between video quality and playback latency by adjusting the system parameter $V$.
	Among comparison techniques, the proposed scheme improves the quality performance sufficiently while minimizing the increase in delay incidence by taking large $V$.
	Quality improvements of other comparison techniques due to large $V$ are comparable to that of the proposed one, but delay performances of `Highest-Quality' and `One-Step' are still much worse than that of the proposed one and `Strongest'. 
	As we've seen in Sections \ref{subsec:simul-node} and \ref{subsec:simul-cachingProb}, the proposed technique provides higher average video quality than `Strongest' and delay performance almost same as `Strongest'.
	We can also see that `One-Step' does not respond sensitively to changes in $V$ compared to other techniques, because the role of $V$ is not completely captured in this scheme. 
	To reflect the effect of $V$ properly, minimization of the frame-based drift-plus-penalty term is necessary, but decisions of `One-Step' on quality and chunk amounts are not frame-based. 
	Those decisions are just conducted and dependent on only each time slot.
	
	\begin{figure}[t]
		\minipage{0.47\textwidth}
		\includegraphics[width=\linewidth]{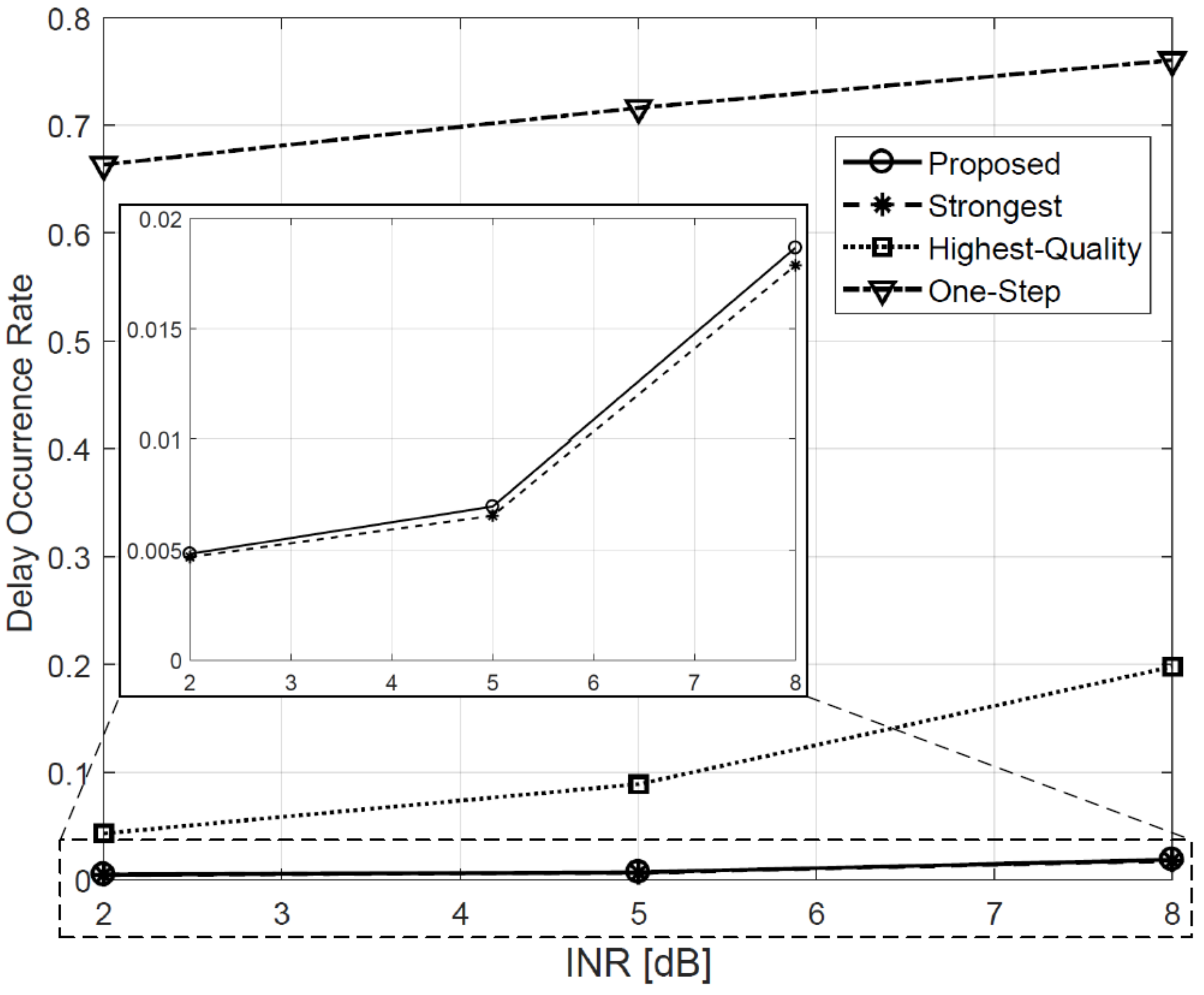}
		\caption{Delay occurrence rates over $\Upsilon$}
		\label{fig:delay_INR}
		\endminipage\hfill
		\minipage{0.47\textwidth}
		\includegraphics[width=\linewidth]{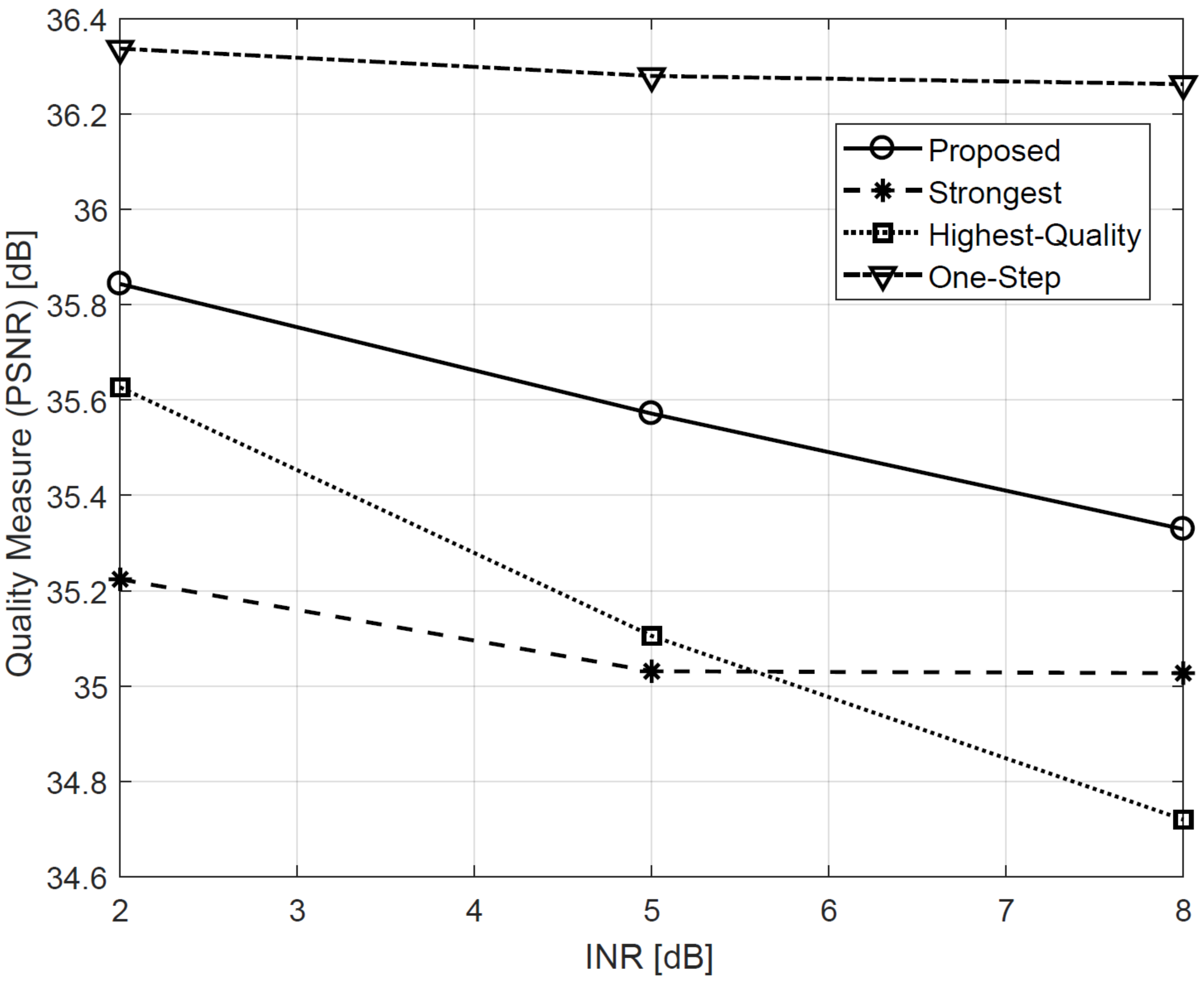}
		\caption{Video quality measures over $\Upsilon$}
		\label{fig:quality_INR}
		\endminipage
	\end{figure}
	
	\subsection{SINR level}
	\label{subsec:simul-SINR}
	
	The delay and quality performances over INR levels are shown in Figs. \ref{fig:delay_INR} and \ref{fig:quality_INR}, respectively.
	It is easily expected that quality performances decrease and delay occurrence rates increase as INR grows for all comparison techniques.
	Almost all of the performance rankings among comparison techniques remain as seen former subsections, but the performance of `Highest-Quality' is influenced by INR levels much more than the proposed one and `Strongest'.
	We can expect that `Highest-Quality' becomes more difficult to accumulate video chunks in the queue as the INR grows, therefore the quality level chosen by the user becomes increasingly degraded.
	%could provide the better video quality even its delay performance is worse than those of the proposed technique and `Strongest'.
	Rather, `Strongest' is not significantly affected by INR changes compared to `Highest-Quality', because the channel condition of its caching node is much stronger than that of the node chosen by `Highest-Quality'.
	The proposed scheme still achieves the improved video quality while guaranteeing very low delay occurrence rate.
	
	\section{Conclusion}
	\label{sec:conclusion}
	
	This paper studies the dynamic delivery policy of video files of various quality levels in the wireless caching network.
	When the caching node distribution around the streaming user is varying, e.g. the user is moving, the streaming user makes decisions on caching node to receive the desired file, video quality, and the number of receiving chunks.
	The different timescales are considered for the caching node association and decisions on quality and the number of receiving chunks.
	The optimization framework of those video delivery decisions conducted on different timescales is constructed based on Lyapunov optimization theory and Markov decision process.
	By using dynamic programming and the frame-based drift-plus-penalty algorithm, the dynamic video delivery policy is proposed to maximize average streaming quality while limiting playback delay quite low.
	Further, the proposed technique can adjust the tradeoff between performances of video quality and playback delay by controlling the system parameter of $V$.
	
	\section*{Acknowledgment}
	This work has supported by Institute for Information \& communications Technology Planning \& Evaluation (IITP) grant funded by the Korea government (MSIT) (No.2018-0-00170, Virtual Presence in Moving Objects through 5G).

	\vspace{12pt}


\begin{thebibliography}{00}
		\bibitem{cisco} ``Cisco Visual Networking Index: Global Mobile Data Traffic Forecast Update, 20162021 White Paper”, Cisco. [Online]. Available: https://www.cisco.com/c/en/us/solutions/collateral/serviceprovider/visual-networking-index-vni/mobile-white-paper-c11-520862.html
		
		\bibitem{youtube}
		X. Cheng, J. Liu, and C. Dale, ``Understanding the characteristics of Internet short video sharing: A YouTube-based measurement study," \textit{IEEE Trans. Multimedia}, 15(5):1184--1194, Aug. 2013.
		
		\bibitem{mm17koo}
		J. Koo, J. Yi, J. Kim, M. A. Hoque, and S. Choi, ``REQUEST: Seamless dynamic adaptive streaming over HTTP for multi-homed smartphone under resource constraints," in \textit{Proc. of ACM Multimedia}, Mountain View, CA, 2017.
		
		\bibitem{femtocaching}
		N. Golrezaei, K. Shanmugam, A. G. Dimakis, A. F. Molisch, and G. Caire, ``FemtoCaching: Wireless video content delivery through distributed caching helpers," in \textit{Proc. of IEEE INFOCOM}, Orlando, FL, USA, 2012.
		
		\bibitem{CM2014Bastug}
		E. Bastug, M. Bennis, and M. Debbah, ``Living on the edge: The role of proactive caching in 5G wireless networks," \textit{IEEE Communications Magazine}, 52(8):82--89, Aug. 2014.
		
		\bibitem{CM2014Wang}
		X. Wang, M. Chen, T. Taleb, A. Ksentini, and V. C. M. Leung, ``Cache in the air: exploiting content caching and delivery techniques for 5G systems," \textit{IEEE Communications Magazine}, 52(2):131--139, Feb. 2014.
		
		\bibitem{caching:ICC2015Blaszczyszyn}
		B. Blaszczyszyn and A. Giovanidis, ``Optimal geographic caching in cellular networks," in  \textit{Proc. of IEEE Int'l Conf. Communi. (ICC)}, London, 2015, pp. 3358--3363.
		
		\bibitem{CL2017Chen}
		Z. Chen, N. Pappas, and M. Kountouris, ``Probabilistic Caching in Wireless D2D Networks: Cache Hit Optimal Versus Throughput Optimal," \textit{IEEE Communications Letters}, vol. 21, no. 3, pp. 584–587, March 2017.
		
		\bibitem{caching:TWC2016Chae}
		S. H. Chae and W. Choi, ``Caching placement in stochastic wireless caching helper networks: Channel selection diversity via caching," \textit{IEEE Trans. Wireless Communi.}, 15(10):6626--6637, Oct. 2016.
		
		\bibitem{caching:TC2016Malak}
		D. Malak, M. Al-Shalash, and J. G. Andrews, ``Optimizing content caching to maximize the density of successful receptions in device-to-device networking," \textit{IEEE Trans. Comm.}, 64(10):4365--4380, Oct. 2016.
		
		\bibitem{caching:ICC2019Choi}
		M. Choi, D. Kim, D.-J. Han, J. Kim, and J. Moon, ``Probabilistic Caching Policy for Categorized Contents and Consecutive User Demands," \textit{IEEE International Conference on Communications (ICC)}, 2019.
		
		\bibitem{JSAC2016Gregori}
		M. Gregori, J. G´omez-Vilardeb´o, J. Matamoros, and D. G¨und¨uz, ``Wireless Content Caching for Small Cell and D2D Networks," \textit{IEEE Journal on Selected Areas in Communications}, vol. 34, no. 5, pp. 1222–1234, May 2016.
		
		\bibitem{TWC2018Emara}
		M. Emara, H. Elsawy, S. Sorour, S. Al-Ghadhban, M. Alouini and T. Y. Al-Naffouri, ``Optimal Caching in 5G Networks With Opportunistic Spectrum Access," \textit{IEEE Transactions on Wireless Communications}, vol. 17, no. 7, pp. 4447-4461, July 2018.
		
		\bibitem{cachingDiffQual:infocom2014Poularakis}
		K. Poularakis, G. Iosifidis, A. Argyriou, and L. Tassiulas, ``Video delivery over heterogeneous cellular networks: Optimizing cost and performance," in \textit{Proc. of IEEE INFOCOM}, Toronto, ON, 2014.
		
		\bibitem{cachingDiffQual:CL2017Zhan}
		C. Zhan and Z. Wen, ``Content cache placement for scalable video in heterogeneous wireless network," \textit{IEEE Communi. Lett.}, 21(12):2714--2717, Dec. 2017.
		
		\bibitem{infocom2016Poularakis}
		K. Poularakis, G. Iosifidis, A. Argyriou, I. Koutsopoulos and L. Tassiulas, ``Caching and operator cooperation policies for layered video content delivery," \textit{IEEE INFOCOM 2016 - The 35th Annual IEEE International Conference on Computer Communications}, San Francisco, CA, 2016, pp. 1-9.
		
		\bibitem{cachingDiffQual:CL2016Wu}
		L. Wu and W. Zhang, ``Caching-based scalable video transmission Over cellular networks," \textit{IEEE Communi. Lett.}, 20(6):1156--1159, Jun. 2016.
		
		\bibitem{JSAC2018Choi} M. Choi, J. Kim and J. Moon, ``Wireless Video Caching and Dynamic Streaming Under Differentiated Quality Requirements," \textit{IEEE Journal on Selected Areas in Communications,} vol. 36, no. 6, pp. 1245-1257, June 2018.
		
		\bibitem{TWC2016Yang}
		C. Yang, Y. Yao, Z. Chen and B. Xia, ``Analysis on Cache-Enabled Wireless Heterogeneous Networks," \textit{IEEE Transactions on Wireless Communications}, vol. 15, no. 1, pp. 131-145, Jan. 2016.
		
		\bibitem{NA:TC2014Poularakis}
		K. Poularakis, G. Iosifidis, and L. Tassiulas, ``Approximation algorithms for mobile data caching in small cell networks," \textit{IEEE Trans. Comm.}, 62(10):3665--3677, Oct. 2014.
		
		\bibitem{NA:TC2016Zhang}
		L. Zhang, M. Xiao, G. Wu, and S. Li, ``Efficient scheduling and power allocation for D2D-assisted wireless caching networks," \textit{IEEE Trans. Communi.}, 64(6):2438--2452, Jun. 2016.
		
		\bibitem{NA:TMC2017Jiang}
		W. Jiang, G. Feng, and S. Qin, ``Optimal cooperative content caching and delivery policy for heterogeneous cellular networks," \textit{IEEE Trans. Mobile Comput.}, 16(5):1382--1393, May 2017.
		
		\bibitem{VD_DiffQ:TM2013Wang}
		X. Wang, M. Chen, T. T. Kwon, L. Yang, and V. C. M. Leung, ``AMESCloud: A framework of adaptive mobile video streaming and efficient social video sharing in the clouds", \textit{IEEE Trans. Multimedia}, 15(4):811--820, Jun. 2013.
		
		\bibitem{VD_DiffQ:TC2015Bethanabhotla}
		D. Bethanabhotla, G. Caire, and M. J. Neely, ``Adaptive video streaming for wireless networks with multiple users and helpers," \textit{IEEE Trans. Comm.}, 63(1):268--285, Jan. 2015.
		
		\bibitem{VD_DiffQ:TON2016Kim}
		J. Kim, G. Caire, and A. F. Molisch, ``Quality-aware streaming and scheduling for device-to-device video delivery," \textit{IEEE/ACM Trans. Netw},  24(4):2319--2331, Aug. 2016. 
		
		\bibitem{TC2018Yang}
		J. Yang, P. Si, Z. Wang, X. Jiang and L. Hanzo, ``Dynamic Resource Allocation and Layer Selection for Scalable Video Streaming in Femtocell Networks: A Twin-Time-Scale Approach," \textit{IEEE Transactions on Communications}, vol. 66, no. 8, pp. 3455--3470, Aug. 2018.
		
		\bibitem{dash}
		T. Stockhammer, ``Dynamic adaptive streaming over HTTP - standards and design principles,” \textit{Proc. ACM MMSys2011}, California, Feb. 2011. 
		
		\bibitem{TAC2013Neely} M. J. Neely and S. Supittayapornpong, ``Dynamic Markov decision policies for delay constrained wireless scheduling," \textit{IEEE Trans. Automatic Control}, 58(8):1948--1961, Aug. 2013.
		
		\bibitem{DynamicProgram} D. P. Bertsekas, \textit{Dynamic Programming and Optimal Control}, 4th Ed., Athena Scientific, 2017
		
		\bibitem{TWC2017Chen}
		Z. Chen, J. Lee, T. Q. S. Quek and M. Kountouris, ``Cooperative Caching and Transmission Design in Cluster-Centric Small Cell Networks," \textit{IEEE Transactions on Wireless Communications}, vol. 16, no. 5, pp. 3401-3415, May 2017.
		
		\bibitem{TWC2018Qin}
		Z. Qin, X. Gan, L. Fu, X. Di, J. Tian and X. Wang, ``Content Delivery in Cache-Enabled Wireless Evolving Social Networks," \textit{IEEE Transactions on Wireless Communications}, vol. 17, no. 10, pp. 6749-6761, Oct. 2018.
		
		\bibitem{DySPAN2005Etkin}
		R. Etkin, A. Parekh and D. Tse, ``Spectrum sharing for unlicensed bands," \textit{IEEE International Symposium on New Frontiers in Dynamic Spectrum Access Networks, 2005. DySPAN 2005.,} Baltimore, MD, USA, 2005, pp. 251-258.
		
		\bibitem{ICC2009Blomer}
		J. Blomer and N. Jindal, 11Transmission Capacity of Wireless Ad Hoc Networks: Successive Interference Cancellation vs. Joint Detection," \textit{2009 IEEE International Conference on Communications}, Dresden, 2009, pp. 1-5.
		
		\bibitem{LittlesThm} D. Bertsekas and R. Gallager, \textit{Data Networks}, 2nd Ed., Prentice, 1992. 
		
		\bibitem{Lyapunov} M. J. Neely, \textit{Stochastic Network Optimization with Application to Communication and Queueing Systems}, Morgan \& Claypool, 2010.
		
		\bibitem{ICTC2015Kim} J. Kim and E. S. Ryu, ``Feasibility study of stochastic streaming with 4K UHD video traces," in \textit{Proc. IEEE Int'l Conf. Information and Communi. Technology Convergence (ICTC)}, 2015, pp. 1350-1355.
	\end{thebibliography}
\end{document}